\documentclass[psfig,txfonts]{aa}
\usepackage{natbib}
\usepackage{psfig}
\bibpunct{(}{)}{;}{a}{}{,}

\newcommand{\kms}{\mbox{\,km\,s$^{-1}$}}
\newcommand{\Kkms}{\mbox{\,K\,km\,s$^{-1}$}}
\newcommand{\Msun}{\,M$_{\odot}$}
\newcommand{\Lsun}{\,L$_{\odot}$}
\newcommand{\HII}{\mbox{$\mathrm{H\,{\scriptstyle {II}}}$}}

\newcommand{\hh}{$\mathrm{H_{2}}$}
\newcommand{\coh}{\mbox{$^{12}${\rmfamily CO}{(2--1)}}}
\newcommand{\col}{\mbox{$^{12}${\rmfamily CO}{(1--0)}}}
\newcommand{\co}{$^{12}${\rmfamily CO}$\,$}
\newcommand{\tco}{\mbox{$^{13}${\rmfamily CO}}}
\newcommand{\tcoh}{\mbox{$^{13}${\rmfamily CO}{(2--1)}}}

\newcommand{\ceol}{\mbox{\rmfamily C}$^{18}${\rmfamily O}{(1--0)}}
\newcommand{\cs}{\mbox{\rmfamily CS}}
\newcommand{\csl}{\mbox{\rmfamily CS}{(2--1)}}
\newcommand{\csm}{\mbox{\rmfamily CS}{(3--2)}}
\newcommand{\um}{$\mu$m}
\newcommand{\nCar}{\mbox{$\eta$ Car}}

\newcommand{\tinner}{T$_{\mathrm {inner}}$}
\newcommand{\rinner}{R$_{\mathrm {inner}}$}
\newcommand{\router}{R$_{\mathrm {outer}}$}
\newcommand{\touter}{T$_{\mathrm {outer}}$}

\newcommand{\twomass}{{2MASS}}
\newcommand{\MSX}{{\it MSX}}
\newcommand{\IRAS}{{\it IRAS}}

\newcommand{\sIII}{\mbox{G287.73-0.92}}
\newcommand{\sIV}{\mbox{G287.84-0.82}}
\newcommand{\sV}{\mbox{G287.93-0.99}}
\newcommand{\sVI}{\mbox{G288.07-0.80}}
\newcommand{\myso}{\mbox{G287.87-1.36}}

\newcommand{\Wmmsr}{\mbox{\,W\,m$^{-2}$\,sr$^{-1}\,$}}
\newcommand{\fdc}{\mbox{F$_{14}$/F$_{12}$}}
\newcommand{\fea}{\mbox{F$_{21}$/F$_{8}$}}

\begin{document}
\title{The giant pillars of the Carina Nebula}
\author{J.M. Rathborne\inst{1,2} \and K.J. Brooks\inst{1,3} \and M.G. Burton\inst{2} \and M. Cohen\inst{4} \and S. Bontemps\inst{5}}
\institute{European Southern Observatory, Casilla 19001, Santiago 19, Chile \and School of Physics, The University of New South Wales, Sydney, NSW, 2052, Australia \and Departamento de Astronomia, Universidad de Chile, Casilla 36-D, Santiago, Chile \and Radio Astronomy Laboratory, 601 Campbell Hall, University of California, Berkeley, CA 94720, USA \and Observatoire de Bordeaux, BP 89, 33270 Floirac, France}
\offprints{J.M. Rathborne \email{jmr@phys.unsw.edu.au}}

\date{Received ... / Accepted ...}


\abstract{
Results are presented from a multi-wavelength study of the giant pillars within 
the Carina Nebula. Using near-IR data from \twomass, mid-IR data from \MSX, 843\,MHz
radio continuum maps from the MOST and molecular line and continuum observations from the SEST, we 
investigate the nature of the pillars and search for evidence of ongoing star formation within them.
Photodissociation regions (PDRs) exist across the whole nebula and trace the giant 
pillars, as well as many ridges, filaments and condensations (A$_{\mathrm v} >$ 7~mag). Morphological 
similarities between emission features at 21\,\um\, and 843\,MHz adjacent
to the PDRs, suggests that the molecular material has been carved by the intense stellar 
winds and UV radiation from the nearby massive stars. In addition, star forming cores are found at the tips of 
several of the pillars. Using a stellar density distribution, several candidate embedded clusters are also found.
One is clearly seen in the \twomass\, images and is located within a dense core (\sIV). 
A search for massive young stellar objects and compact \HII\, regions using mid-IR colour criteria, reveal
twelve candidates across the complex. Grey-body fits to SEDs for four of these objects are suggestive of OB-stars. 
We find that massive star formation in the Carina Nebula is occurring across the whole complex and 
confirm it has been continuous over the past 3~Myrs.

}

\authorrunning{Rathborne et al.}
\maketitle

\section{Introduction}

In many cases, young stars have been found at the tips of giant pillars that
point toward a more evolved massive star cluster (e.g. the elephant trunks of the Eagle
Nebula; \citeauthor{McCaughrean02}\, \citeyear{McCaughrean02}). The formation of such pillars can readily occur if a dense 
core within a giant molecular cloud (GMC) is exposed to the intense stellar winds and radiation 
fields from a nearby massive star cluster. The core would shield the column of molecular material 
behind it, in a direction pointing away from the cluster. Subsequently, the more exposed parts of 
the GMC would be swept up around this column or be completely irradiated away. 

It is not clear if such a drastic change in the structure of a GMC can affect its star formation 
capacity. There is growing evidence to suggest that the tips of pillars are prime sites for ongoing 
star formation \citep{Jiang02, Stanke02, McCaughrean02}. However, there is much 
debate over whether this type of star formation has been triggered by external processes, or whether 
it has spontaneously formed. It is uncertain if we can even distinguish between the two.

Several giant pillars pointing toward the massive clusters within the Carina Nebula have recently
been discovered at mid-infrared (mid-IR) wavelengths, the largest extending $\sim$25pc \citep{Smith00}. Bright 
IR emission condensations are located at the tips of several of these pillars, and may correspond to sites 
where star formation has been triggered by the interactions with the surrounding young massive star 
clusters \citep{Smith00}.

Many large molecular clouds are associated with the Carina Nebula \citep{Zhang01,
Brooks98, Whiteoak84, deGraauw81}. These clouds lie close to several 
massive star clusters, in particular Bochum 10 and 11, Collinder 228 and Trumpler 14, 15 and 16 
(hereafter the cluster abbreviations Bo, Co and Tr will be used). These clusters contain a combined 
total of 64 O-type stars including one of the most massive and spectacular stars known, \nCar\, 
\citep{Feinstein95}. Located at a distance of 2.2~kpc \citep{Tovmassian95}, the Carina Nebula is an 
excellent region in which to study the effect massive stars have on their natal GMC. 

The giant pillars are located in the relatively unstudied southern region of this 
nebula, at a greater distance from the most influential clusters. The stellar winds
and radiation fields may be less destructive here, making the pillars prospective sites for ongoing 
star formation. It is the aim of this paper to investigate the nature of the interstellar medium 
within, and surrounding these pillars, and in particular to determine if there exists any evidence for 
ongoing star formation within them.

\section{The data}
\label{data}

\subsection{\MSX\, on-line database}
\label{msx-data}

Images between 8--21\,\um\, were obtained from the 
\MSX\, database\footnote{For a full description of the \MSX~satellite, the astrophysical experiments 
and the observing techniques see \citet{Mill94, Price96, Egan98, Price01}.}, and cover four discrete bands; 
Band A (6.8--10.8\,\um), 
{\mbox Band C} (11.1--13.2\,\um), Band D (13.5--15.9\,\um) and Band E (18.2--25.1\,\um). These bands will 
be hereafter referred to as the 8, 12, 14, and 21\,\um\, bands respectively. The angular resolution
of the images (FWHM) is 20\arcsec\, with an astrometric accuracy of $\sim$ 2\arcsec.

Over the four \MSX\, bands, the emission arises from either fluorescently excited polycyclic 
aromatic hydrocarbon (PAH) molecules or black-body emission from heated dust grains (150--400\,K). Emission 
from the family of PAH features 
(6.2, 7.7, 8.7, 11.3, 12.7 and 16.4\,\um), as well as several emission plateaus caused by small
PAH molecules (6-9\,\um, 11-13\,\um\, and 15-20\,\um) are all contained within the 8, 12 and 14\,\um\, 
bands \citep{Verstraete01}. The emission seen in the {\mbox 21\,\um\,} band generally arises from heated
dust, and is useful in revealing the locations of heavily obscured young stars. 

The \MSX\, Point Source Catalogue (PSCv1.0) provided position and flux
measurements for point sources within the Carina Nebula (over the extent shown in Fig.~\ref{8um-Av}). 
A total of 526 point sources were found, however, the majority are only detected in the
higher sensitivity 8\,\um\, band. For the discussion presented in Sect.~\ref{myso-candidates}, a subset of this 
larger catalogue was used. Sources were extracted from the PSC if they had reliable fluxes in each of the 8-, 14- 
and 21\,\um\, bands (corresponding to quality flags 3 or 4). This reduced the list to a total 
of 33 sources. 

\subsection{MOST on-line database}
\label{most-data}

The radio continuum data presented here were obtained as part of the Molonglo Galactic Plane 
Survey (MGPS) made with the Molonglo Observatory Synthesis Telescope 
(MOST)\footnote{http://www.physics.usyd.edu.au/astrop/most/}. The centre frequency of the MOST is 
843\,MHz, with a bandwidth of 3\,MHz. The synthesized beam is 
{\mbox {43\arcsec$\times$43\arcsec cosec$|\delta|$}}. For more details of the 
specifications of the MOST see \citet{Mills81} and \citet{Robertson91}.

\subsection{SEST observations}
\label{sest-data}

\subsubsection{Spectral line observations}
Spectral line observations were obtained toward four mid-IR condensations 
(\sIII, \sIV, \sV\, and \sVI; \citeauthor{Smith00} \citeyear{Smith00}) using the \mbox{15-m} Swedish-ESO 
Submillimetre Telescope (SEST)\footnote{The SEST is operated jointly by ESO and the Swedish National Facility for Radio Astronomy, Onsala Space Observatory, Chalmers University of Technology.} at La Silla Observatory, Chile. The 
condensations were observed in six molecular lines: \col, \coh, \tcoh, \ceol, \csm\, and \csl. Details of the 
observations are given in Table~\ref{obs-parameters}. 

Both the IRAM-built 230/115\,GHz and the SESIS 150/100\,GHz receivers were used. 
Transitions were observed simultaneously where possible in single side band mode (SSB), with the 
receivers connected to a narrow-band (43\,MHz) Acousto-Optical Spectrometer (AOS). 

Position-switching mode was used for all observations, with \nCar\, used as the signal-free 
reference position.  An off-source position was taken after every second on-source integration. 
Chopper wheel calibration was performed every 10\,mins, to obtain atmosphere-corrected antenna 
temperatures. The telescope pointing and sub-reflector focusing were checked regularly using 
suitably bright nearby SiO masers. We estimate the pointing accuracy to be better than 10\arcsec\,
and adopt the standard SEST value of 10 per cent uncertainty in the temperature scale. 

Observations toward  the condensations were obtained using a 20\arcsec\, pointing grid and 
intended to cover the peak of the emission. However, in several cases data were only obtained
over the tip of much larger structures.

All spectra were processed using GILDAS software \citep{Buisson99}. Baselines were initially 
removed from each spectrum and the temperature scale converted to main-beam brightness temperature 
using the values for the main beam efficiencies ($\eta_{\mathrm mb}$) given in 
Table~\ref{obs-parameters}. For comparison between data sets with differing beam sizes, the data 
were smoothed to the larger beam-size on a similarly sized grid. The beam size of the SEST at 
each frequency, the average system temperatures during the observations, and the average rms 
noise per spectral channel are also given in Table~\ref{obs-parameters}.

\begin{table*}
\begin{center}
\caption{\label{obs-parameters}Observing parameters for the molecular line data obtained with the SEST. t$_{\mathrm int}$ refers to the integration time, HPBW is the half-power beam-width, $\eta_{\mathrm mb}$ is the main beam efficiency, T$_{\mathrm sys}$ is the average system temperature, and $\sigma$ is the average rms noise per spectral channel for the observations.}
\begin{minipage}{0.7\textwidth}
\begin{center}
\begin{tabular}{rccccccc}
\hline
Transition&Frequency&Date & t$_{\mathrm int}$ & HPBW & $\eta_{\mathrm mb}$  & T$_{\mathrm sys}$ & $\sigma$ \\
       &     GHz   &      &s          &{\footnotesize arcsec}  & &K & K   \\
\hline
\col   & 115.217   & Oct 2000       & 30  & 45 & 0.70 & 529 & 0.7\\
\coh   & 230.537   & Oct 2000       & 30  & 23 & 0.50 & 625 & 0.7\\
\ceol  & 109.782   & Mar 2002$^{a}$ & 120 & 49 & 0.72 & 321 & 0.4\\
\tcoh  & 220.398   & Mar 2002$^{a}$ & 120 & 24 & 0.52 & 488 & 0.5\\
\csl   &  97.981   & Jan 2002       & 60  & 52 & 0.73 & 337 & 0.4\\
\csm   & 146.969   & Jan 2002       & 60  & 34 & 0.66 & 433 & 0.5\\
\hline
\end{tabular}\\
\begin{minipage}{15cm}
$^{a}$ The observations at these frequencies for source \sV\, were obtained in Oct 2000.
\end{minipage}
\end{center}
\end{minipage}
\end{center}
\end{table*}

\subsubsection{Continuum observations}

Observations of 1.2-mm continuum emission were also obtained with the SEST using the 37 channel 
Imaging Bolometer Array (SIMBA). The FWHM of each element is 23\arcsec\, within the array, with a separation between
each of 44\arcsec. The observations were conducted in July 2002 in the fast-scanning 
mode (80 arcsec s$^{-1}$). Data for a single point source in the Carina Nebula (\myso) were obtained 
using a map size of {\mbox {480\arcsec\, $\times$ 240\arcsec\,}} (az $\times$ elev). To complete this 
map, 31 sub-scans were required with each sub-scan separated by 8\arcsec. Each map took 
$\sim$ 4 mins with the map repeated 6 times to improve the signal to noise. The pointing and 
sub-reflector focusing were checked prior to the observations with sky-dip calibrations performed 
both before and after the mapping. Observations of Uranus were also obtained for flux calibration. 
The average opacity during the observations was {\mbox {$\sim$ 0.14}}.

All maps were reduced by applying the opacity corrections, fitting and subtracting a 
baseline, and removing the correlated sky noise. Maps were flux calibrated using the conversion factor
obtained from the Uranus observations. The rms noise level in the final map was
9~mJy. All data reduction was achieved within the MOPSI package\footnote{MOPSI was developed by R. Zylka (Astrophysikalisches Institut, Universit\"at zu K\"oln, Germany).}.

\subsection{\twomass\, on-line database}
\label{2mass-data}

Near-IR images  were obtained from the Two Micron All Sky Survey 
(\twomass)\footnote{see http://www.ipac.caltech.edu/2mass/index.html} 
Second Incremental Data Release. The images cover the three near-IR bands: 
J~(1.24\,\um), H~(1.66\,\um), and K$_{\mathrm s}$~(2.16\,\um) down to limiting magnitudes of 15.8, 15.1 
and 14.3 respectively. The resolution of the final images is 2\arcsec. 

\section{The nature of the pillars}
\label{pillars}

This section aims to identify, and determine the nature of, the pillars within the southern Carina
Nebula. We will attempt to trace the extent of the gas and dust within the region, determine how
the radiation from the nearby clusters is penetrating through the GMC, locate ionization
fronts, and attempt to characterise the properties of regions where star formation may be occurring.

\begin{figure*}
\psfig{file=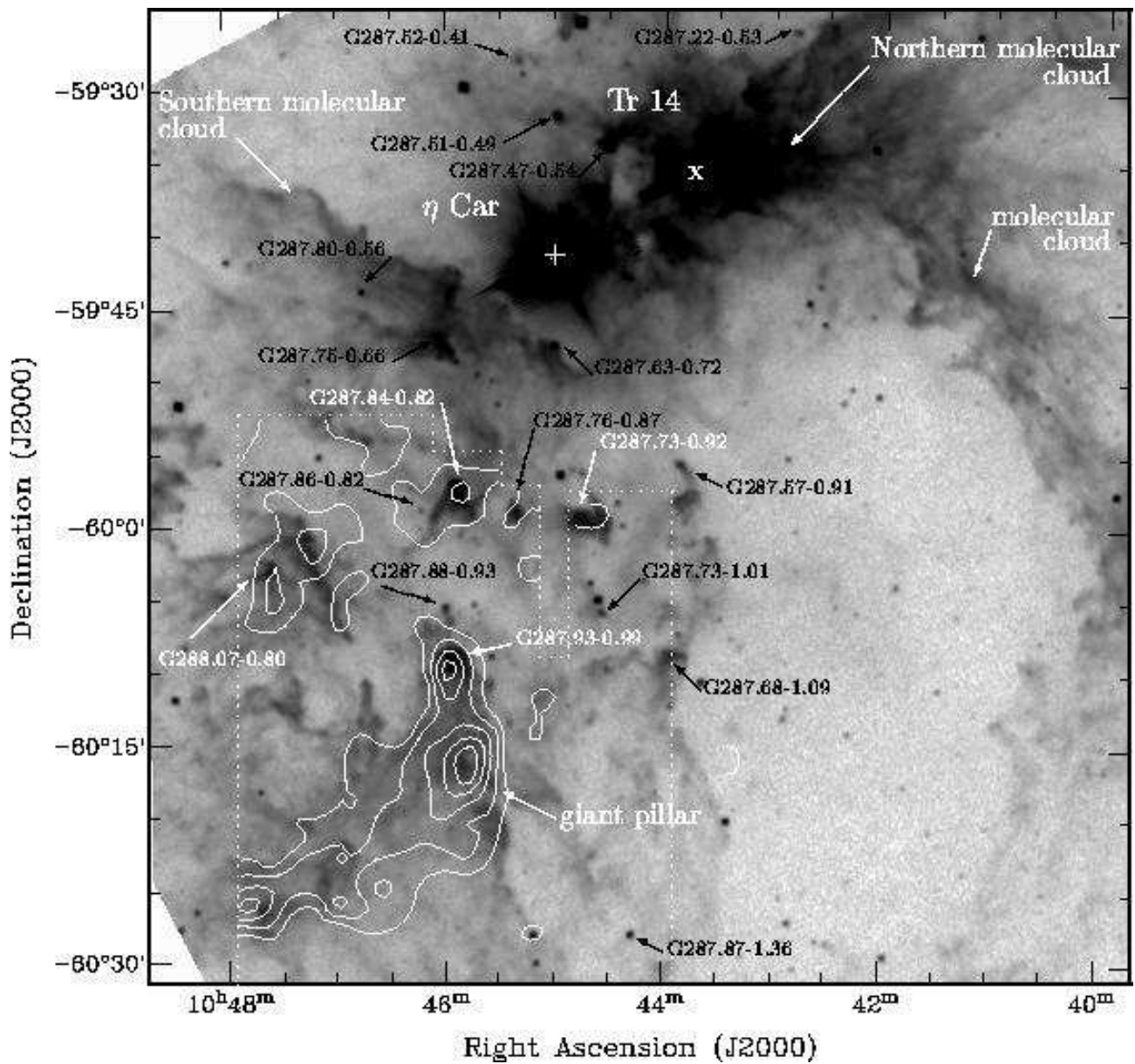,width=0.95\textwidth}
\caption{\label{8um-Av}The Carina Nebula  in 8\,\um\, emission (logarithmic grey-scale) with 
contours of visual extinction (levels are 7.5, 9.5, 11.5, 13.5 and 15.5~mag). Marked on the figure 
are the locations of \nCar\, (+), Tr~14 (x), the southern and northern molecular clouds, and the 
additional giant pillar and molecular cloud identified by Zhang et al. (2001). The dotted region 
outlines the extent of the extinction map. The four condensations labelled (in white) \sIII, \sIV, 
\sV\, and \sVI\, are discussed in Sect.~\ref{condensations}. The point sources labelled in black 
mark the sources discussed in Sect.~\ref{star-formation}.} 
\end{figure*}

\begin{figure*}
\psfig{file=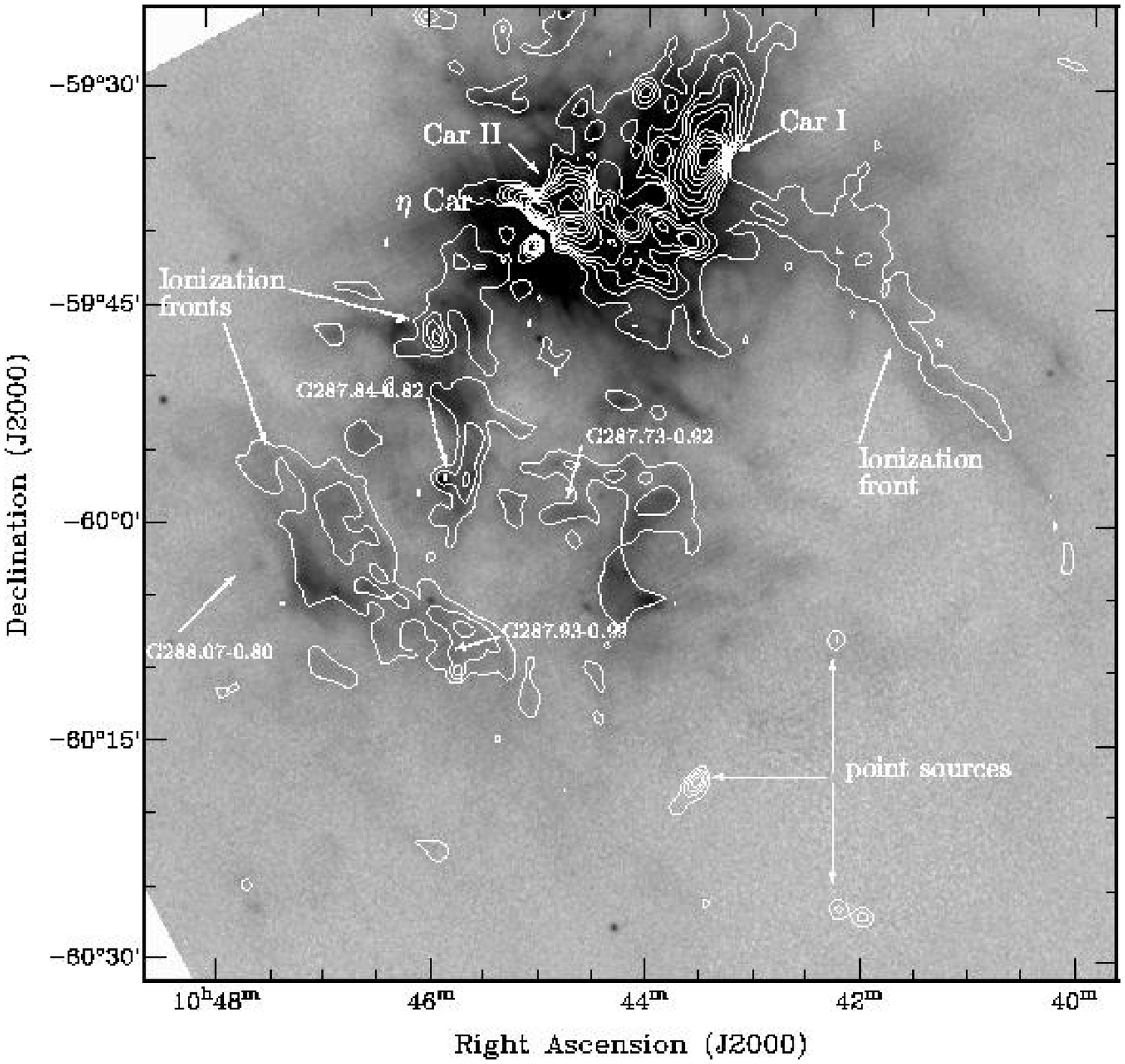,width=0.95\textwidth}
\caption{\label{21um-most}The Carina Nebula in 21\,\um\, emission (logarithmic grey-scale) with
contours of 843\,MHz radio continuum emission (levels are 0.04, 0.14, 0.24, 0.34, 0.5, 0.7, 0.9, 1.1, 
1.3, 1.5 Jy/beam). Marked on this figure is the location of \nCar\, as well as several  
ionization fronts (including Car~I and Car~II), and the four condensations; \sIII, \sIV, \sV\, and 
\sVI.}
\end{figure*}

\subsection{Photodissociation Regions}

Fig.~\ref{8um-Av} shows the Carina Nebula in 8\,\um\, emission.
This image reveals many regions of extended 8\,\um\, emission in addition to several bright 
isolated features. This band is dominated by PAH  
emission and is tracing photodissociation regions (PDRs). These occur within a visual extinction
(A$_{\mathrm v}$) $\sim$1~mag on the surface of molecular material, where the UV radiation has penetrated and is 
heating the material \citep{Hollenbach99}. 

Evident to the northwest of this nebula is a prominent 8\,\um\, feature which corresponds to a component
of the GMC known as the northern molecular cloud. The bright emission south of \nCar\, traces 
the edge of the southern molecular cloud. Located further to the south is the giant pillar
identified by \citet{Smith00}. These clouds are all identified in the large scale {\mbox{CO(4--3)}} 
and [CI] survey of \citet{Zhang01}. These transitions are tracers of heated gas ($\sim$50K), dense 
gas (n $\sim$10$^{5}$~cm$^{-3}$), and PDRs, respectively, and confirm that 
the 8\,\um\, emission is tracing the outer PDR layers of the molecular clouds.

Fig.~\ref{8um-Av} also illustrates the mid-IR sources identified by \citeauthor{Smith00} (2000;
\sIII, \sIV, \sV\, and \sVI). Many other point sources are also
found across the complex, and are located at the tips of more extended 8\,\um\,
emission (e.g.\,G287.57-0.91 and G287.88-0.93). Their morphology is consistent with a 
dense core being irradiated by Tr~14 and Tr~16.

\subsection{Visual extinction}
\label{extinction}

Extinction maps were constructed for the southern Carina Nebula using stars within the \twomass\, 
PSC. Estimates of the visual extinction were obtained for each star detected in the 
{\mbox{JHK$_{\mathrm s}$-bands}}, by dereddening each to an average main sequence colour in the (J-H) vs.
(H-K$_{\mathrm s}$) plane. The visual extinctions obtained were then used to construct a global 
extinction map for the region (contours on Fig.~\ref{8um-Av}, the extent of the  map is 
marked with the dotted line).

Clearly seen as a region of high visual extinction is the giant pillar, where the morphology of the 
extinction matches the 8\,\um\, emission extremely well. In addition to this giant pillar, other 
regions of high visual extinction  are also found to correspond to 8\,\um\, features. 

The  visual extinction across the giant pillar ranges from 7--15~mag and can be used to estimate the \hh\, 
column density within it. Applying the relation, {\mbox {N(\tco)$\simeq$ 2.7$\times$10$^{15}$ 
A$_{\mathrm v}$}} (cm$^{-2}$; \citeauthor{Frerking82}\,\citeyear{Frerking82}) and converting to N(\hh) using the 
constant of proportionality of 4$\times$10$^{5}$ \citep{Lada94},
the \hh\, column density was found to range between 9--19$\times$10$^{21}$~cm$^{-2}$.

\subsection{Heated dust} 

Fig.~\ref{21um-most} shows the 21\,\um\, and 843\,MHz emission across the Carina Nebula. Of note, is the 
striking morphological similarities of these emission features across the complex.

Strong 21\,\um\, emission surrounds \nCar\, and Tr~14
with diffuse emission found in the south. Several bright point sources are also evident, in particular 
the bright source associated with the {\mbox{8\,\um\,}} condensation \sIV. As mentioned previously, 
extended emission seen within the 21\,\um\, band generally arises from heated dust, while 21\,\um\, point sources 
identify deeply embedded stars. By matching the expected surface brightness of 
an optically thick black-body with the observed flux from regions showing strong 21\,\um\, emission, 
we find a dust temperature of 42--44~K. This temperature is in good agreement with other 
estimates (40K) obtained from \IRAS\, data in the Tr~14 region \citep{Harvey79}.


\subsection{Ionization fronts}

\begin{figure*}[t]
\psfig{file=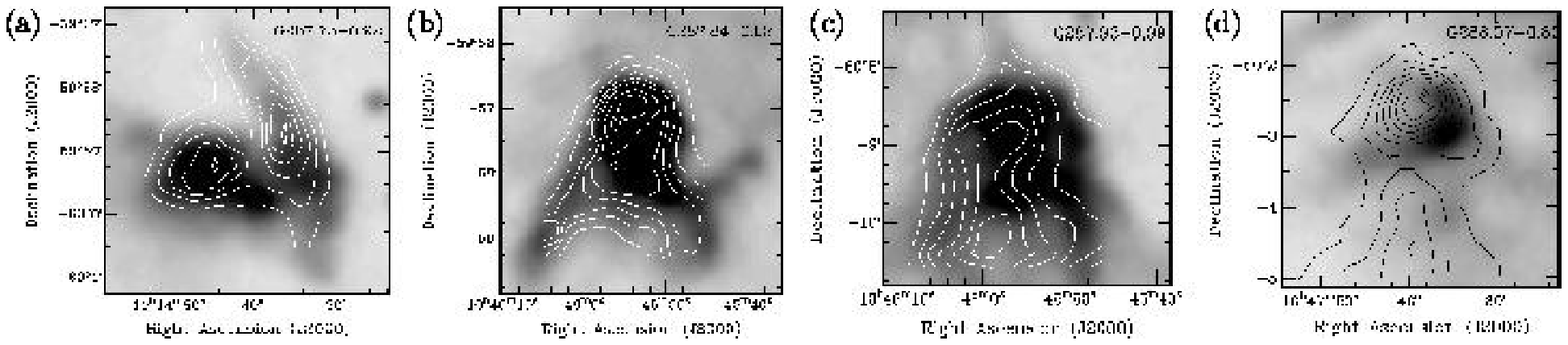,width=0.95\textwidth}
\caption{\label{8um-co}8\,\um\, emission in grey-scale with contours of integrated \coh\, emission for the condensations. (a) \sIII: Integrated from -25 to -32\,\kms, levels are 20, 35, 50, 65, 80, 95\Kkms; (b) \sIV: Integrated from -12 to -16\,\kms, levels are 20, 40, 60, 80, 100, 140, 180\Kkms; (c) \sV: Integrated from -18 to -22\,\kms, levels are 30, 50, 70, 90, 110, 130\Kkms; (d) \sVI: Integrated from -19 to -24\,\kms, levels are 15, 30, 45, 60, 75, 90, 105\Kkms.}
\end{figure*}

Radio continuum emission is widespread throughout the Carina Nebula, with the brightest concentrations 
known as Car~I and Car~II \citep{Brooks01, Whiteoak94, Retallack83, Gardner68}. 

Extending from Car~I is a long filament of weaker emission. Similar features are also seen in the
southern part of the nebula, and are labelled on Fig.~\ref{21um-most}.
All of these features are coincident with 21\,\um\, emission, and in the majority of cases, are
found adjacent to regions of bright 8\,\um\, emission and peaks in the visual extinction map. This 
morphology is consistent with the ionization fronts heating the dust and creating the widespread PDRs.

Compact sources are also evident across the complex.
The majority of these have no mid-IR counterparts (marked as `point sources' on Fig.~\ref{21um-most}),
suggesting they correspond to unresolved background galaxies \citep{Cohen01}. The exception is  
\sIV, which has an extremely strong mid-IR  counterpart. This object is likely to be a compact \HII\, region.

\begin{table*}
\caption{\label{properties}{Properties of the spatially integrated profiles seen toward the condensations. These were obtained from Gaussian fits to the profiles shown in Fig.~\ref{profiles}. The parameters include; the central velocity (V, \kms), peak temperature (T, K) and line-width ($\Delta$V, \kms).}}
\begin{tabular}{lcccccccccccccccccc}
\hline
&\multicolumn{3}{c}{\col} &\multicolumn{3}{c}{\coh}&\multicolumn{3}{c}{\ceol} &\multicolumn{3}{c}{\tcoh}&\multicolumn{3}{c}{\csl} &\multicolumn{3}{c}{\csm}\\
& V & T & $\Delta$V & V & T & $\Delta$V &V & T & $\Delta$V & V & T & $\Delta$V &V & T & $\Delta$V & V & T & $\Delta$V \\
\hline
\multicolumn{10}{l}{\bf{\sIII}} \\
&-31.5&1.0&2.1&-31.5&1.7&1.8&      &   &   &-30.3 &0.4 &1.5 &&&&&&\\
&-27.8&12.8&3.2&-27.9&16.4&3.4&-28.4&0.2&2.2&-28.1& 6.0&1.7&-27.9&0.6&2.3&-27.9&0.6&2.5\\
&     &   &   &     &    &   &     &   &   &-26.7&3.2&2.1 &-25.8&0.2&1.2&-26.0&0.2&0.8\\  
\multicolumn{10}{l}{\bf{\sIV}}  \\
&-14.2&23.8&3.2&-14.2&34.1&3.6&-14.7&0.6&1.9&-14.4&12.0&2.7&-14.4&1.2&2.5&-14.3&1.1&2.4\\
\multicolumn{10}{l}{\bf{\sV}}  \\
&-22.1&6.2&2.0&-22.4&15.3&2.1&-22.6&0.2&1.6&-23.1& 5.0&1.6&-22.7&0.3&2.2&-22.9&0.2&2.4\\
&-19.8&11.5&2.3&-20.3&23.6&2.4&-20.4&1.2&1.4&-20.6&11.0&2.0&-20.4&0.5&1.8&-20.4&0.3&1.7\\
\multicolumn{10}{l}{\bf{\sVI}}  \\
&-21.6&13.0&2.6&-21.7&15.5&2.7&-20.9&0.1&2.2&-21.5& 4.7&2.2&-21.7&0.8&1.9&-21.6&0.7&2.1\\
&-19.3&5.4&2.2&-19.4&6.6&2.3&&&&-19.5&1.5&1.9&-19.6&0.1&1.2&&&\\
\hline
\end{tabular}
\end{table*}
\normalsize

\begin{table*}
\begin{centering}
\caption{\label{ratio-masses}{Physical parameters for the condensations derived from the properties given in Table~\ref{properties}. Listed are: the excitation temperature ($^{12}$T$_{\mathrm ex}$) measured from the \coh\, peak temperature; the ratio of \coh\, and (1--0) integrated intensities ($^{12}$R); the ratio of \tcoh\, and \coh\, integrated intensities ($^{12,13}$R); the ratio of \csm\, and \csl\, integrated intensities ($^{CS}$R); \tcoh\, optical depth ($^{13}\tau$); \tcoh\, column density (N(\tco)); \hh\, column density (N(\hh)); mass estimated under LTE conditions (M$_{\mathrm LTE}$); Virial mass estimate (M$_{\mathrm virial}$) and the average \hh\, density (n(\hh)).}}
\begin{tabular}{lccccccrcccc}
\hline
Condensation&Component & $^{12}$T$_{\mathrm ex}$ & $^{12}$R & $^{12,13}$R & $^{CS}$R & $^{13}\tau$ & N(\tco) & N(\hh) &M$_{\mathrm LTE}$ &M$_{\mathrm virial}$ & n(\hh)\\
             & \kms     & K &&&&&\multicolumn{1}{c}{\footnotesize cm$^{-2}$}&{\footnotesize cm$^{-2}$} &{\footnotesize \Msun}&{\footnotesize \Msun}&{\footnotesize cm$^{-3}$}\\       
\hline
\sIII & -27.8 & 21.7 & 1.34 & 5.43 & 1.00 & 0.20 & 7.0 $\times 10^{15}$ & 2.8 $\times 10^{21}$ & 11  & 84 & 4.1 $\times 10^{3}$ \\
\sIV & -14.2 & 39.5 & 1.64 & 3.88 & 0.89 & 0.40 & 21.5 $\times 10^{15}$ & 8.6 $\times 10^{21}$ &  34 & 197 & 12.5 $\times 10^{3}$\\
\sV & -19.8 & 29.0 & 2.15 & 2.64 & 0.58 & 0.37 & 14.6 $\times 10^{15}$ & 5.7 $\times 10^{21}$ & 45   & 145 & 16.6 $\times 10^{3}$ \\
\sVI & -21.6 & 20.7 & 1.26 & 4.07 & 0.96 & 0.16 & 7.0 $\times 10^{15}$ & 2.8 $\times 10^{21}$ &  11 & 134 & 4.1 $\times 10^{3}$ \\
\hline
\end{tabular}
\end{centering}
\end{table*}
\normalsize

\begin{figure}
\hspace{0.4cm}
\psfig{file=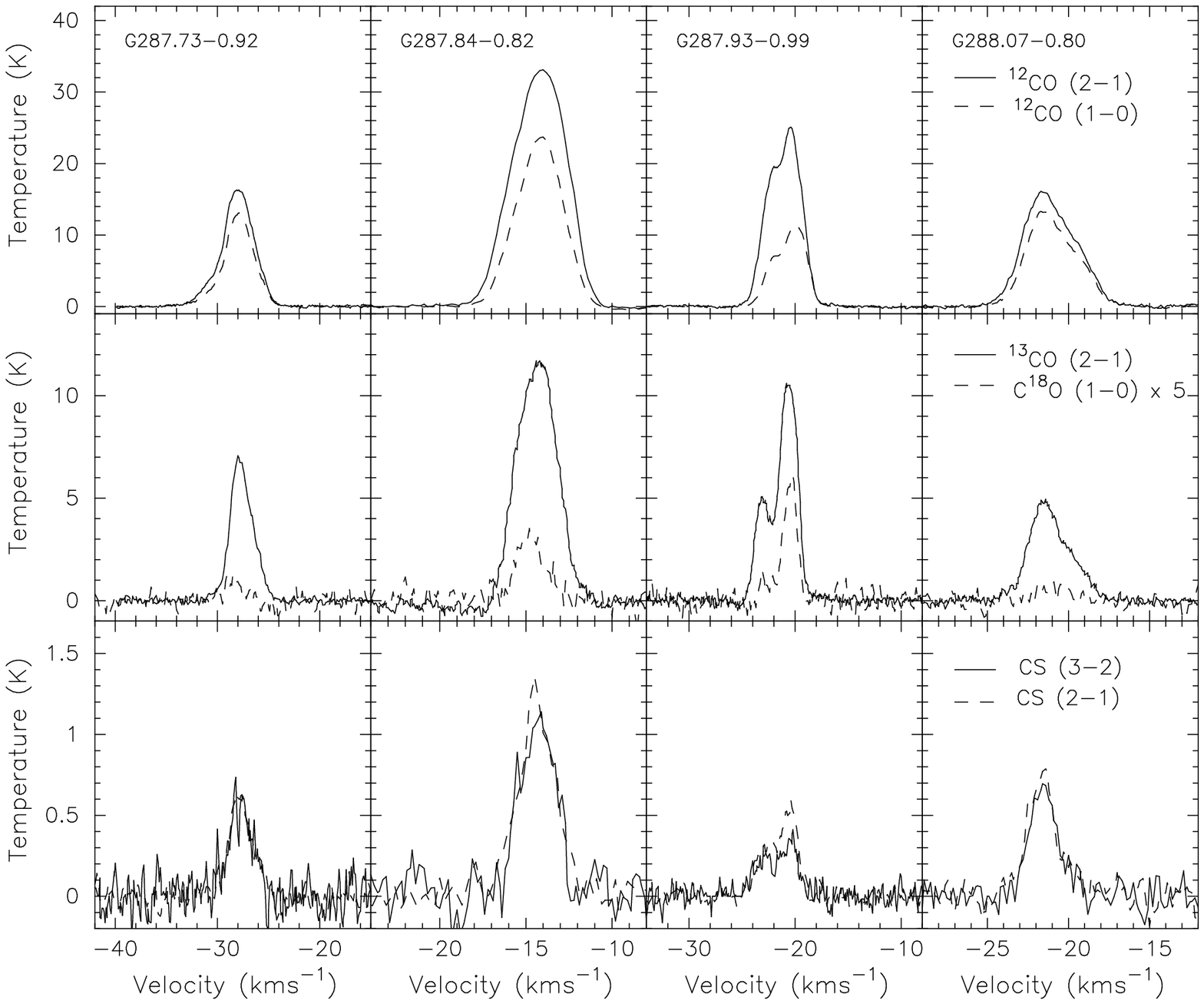,angle=-90,width=0.45\textwidth}
\caption{\label{profiles}Line profiles spatially integrated over each condensation. {\it{Left to right}}: \sIII, \sIV, \sV\, and \sVI. {\it{Top to bottom}}: \coh\, and \col; \tcoh\, and \ceol; \mbox{\csm\,} and \csl\, emission. The \ceol\, profiles are multiplied by 5 to make them visible. Note the different temperature scales and velocity ranges.}
\end{figure}

\subsection{Properties of the four bright mid-IR condensations}
\label{condensations}

Molecular material was detected toward each of the four mid-IR condensations, \sIII, \sIV, \sV\, and 
\sVI. Their integrated \coh\, intensity maps, show the morphology of the molecular 
structures match well with the corresponding 8\,\um\, emission (Fig.~\ref{8um-co}), 
further confirming the presence of PDRs on their surfaces. For \sVI\, however, the geometry of the molecular material 
differs considerably from the 8\,\um\, emission. While the morphology of the molecular gas is consistent with 
irradiation from the direction of the clusters Tr~14 and Tr~16, the PDR appears to point in a different direction. 
Interestingly, this corresponds to the direction of the Bo~11 cluster.

Fig.~\ref{profiles} shows the line profiles for each condensation, in each of the 
molecular transitions observed. The peak temperature (T), central 
velocity (V) and line-width ($\Delta$V) for these profiles were determined from Gaussian fits, the 
results of which are given in Table~\ref{properties}. 

The molecular condensations show a range in velocities of $\sim$15\kms. This 
is consistent with the velocity dispersion of molecular clumps, and motion of the ionized gas, 
observed within the Keyhole Nebula ($\sim$20\kms; \citeauthor{Cox951} \citeyear{Cox951}). This 
suggests there is also a strong interaction between the winds and radiation from the stellar clusters and these 
molecular condensations.

Estimates of the physical properties of the condensations were determined from the 
profiles under the assumption of local thermal equilibrium (LTE). Values for the excitation 
temperatures, line ratios, opacities, masses, and average hydrogen densities for each condensations are 
listed in Table~\ref{ratio-masses}. The ratio of the integrated intensity of the \coh\, and (1--0)
emission ranged between 1--2, and is consistent with optically thick \co\, emission. In addition, excitation 
temperatures were found to range between {\mbox {20--40\,K.}}

Ratios of the \coh\, and \tcoh\, emission ranged between 2--5, suggesting 
that the \tcoh\, transition is optically thin. To estimate the opacity and column density of 
this material, a constant excitation temperature of 35\,K was used. Opacities ranged from 
0.16--0.4, confirming the emission is optically thin. \hh\, column densities were determined to be 
{\mbox {7--22$\times 10^{21}$ cm$^{-2}$}}, and are consistent with values determined when using the 
estimated A$_{\mathrm v}$ derived from the star colours (9--19$\times$10$^{21}$ cm$^{-2}$).

Estimates of the masses were calculated using the column densities determined from the \tco\, 
emission (LTE mass). This calculation assumes the integrated intensity of the emission is 
proportional to the H$_{2}$ column density with a constant of proportionality of 4$\times$10$^{5}$ 
\citep{Lada94}. Masses ranged from 10--45\Msun\, 
for the condensations, however, these may be under-estimated by as much as a factor of 3. 
The Virial mass was estimated using the line-width of the \tco\, line, which is assumed to be a 
measure of the overall motion of the gas and hence the mass. Values of the Virial mass for the 
condensations ranged between 80--200\Msun. These are significantly 
greater than the LTE masses (even considering the latter may be underestimating the true masses), 
implying that the condensations are probably not gravitationally bound but instead confined by 
external pressure. 

The average \hh\, density was determined assuming the size  along the line-of-sight was the average 
of the extent  in right ascension and declination. Estimates ranged 
between 4--17 $\times$ 10$^{3}$ cm$^{-3}$ and represent lower limits to the actual values. In 
addition, the presence of \cs\, emission suggests the material has gas densities in excess of  
10$^{4}$~cm$^{-3}$. The properties described here are consistent with molecular cores 
found in other regions of massive star formation (Zinchenko et al 1998).

\begin{figure}
\begin{center}
\psfig{file=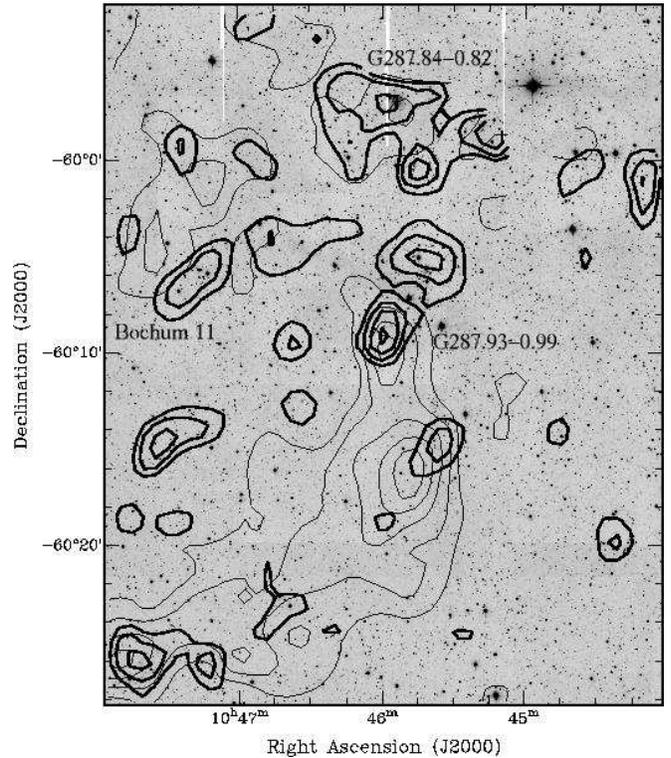,width=0.48\textwidth}
\caption{\label{clusters} K$_{\mathrm s}$-band image for the southern Carina Nebula with contours
representing the stellar density distribution for stars with K$_{\mathrm corr} <$ 12~mag (thick black: levels are 2.6, 
3, 3.4, 3.8~stars/arcmin$^{2}$) and the visual extinction map (thin black: levels are 7.5, 9.5, 11.5, 13.5 and 
15.5~mag). The cluster Bo~11 is marked, along with the condensations \sIV\, and \sV.}
\end{center}
\end{figure}

\section{Star formation within the pillars}
\label{star-formation}

This section investigates the star formation activity across the Carina Nebula, by indentifying
candidate embedded clusters and young stellar objects (YSOs).

\subsection{Young clusters}
 \label{stellar-density}

All stars within the \twomass\, PSC were dereddened. As previously discussed, those detected in the
JH and K$_{\mathrm s}$-bands were dereddened using an average main sequence colour. For those stars only detected
in the H- and/or K$_{\mathrm s}$-bands, the extinction was estimated using nearby stars in the global extinction map. 
From these dereddened K$_{\mathrm s}$-band magnitudes (K$_{\mathrm corr}$) a stellar number density map was created for 
stars with K$_{\mathrm corr} > $12~mag. Choosing this cut-off corresponds to the completeness limit from the \twomass\, 
photometry for an A$_{\mathrm v}$ of~25~mag.

With the foreground extinction removed, the brightest sources will correspond to the 
most massive and youngest objects. Young sources display a near-IR excess, making them brighter in 
the K$_{\mathrm s}$-band than main-sequence stars whose emission is purely from a stellar photosphere.
As stars tend to form in clusters, any over-densities within the stellar density distribution 
will potentially reveal the location of clusters. Fig.~\ref{clusters} displays, as contours, the derived stellar 
density, and visual extinction maps for the southern Carina Nebula. Many over-densities 
are seen here and are candidate young clusters.

\begin{figure}
\begin{center}
\psfig{file=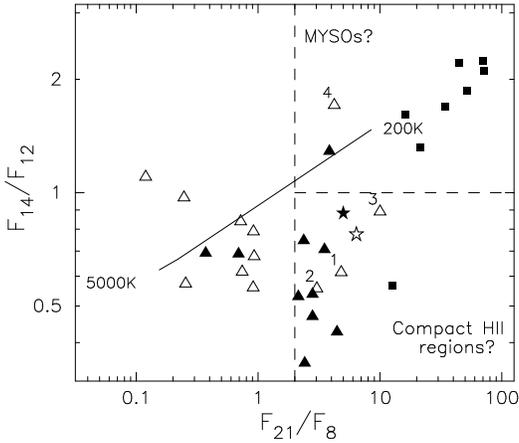,angle=-90,width=0.4\textwidth}
\caption{\label{cc-msx}Colour-colour plot for sources in the Carina Nebula. 
Included on this are the location of a black-body (solid line) and the limits derived by 
Lumsden et al. (2002) for MYSOs and compact \HII\, regions (dashed lines). Plotted here are:
point sources with \MSX\, data (filled triangles), point sources with \MSX\, and \IRAS\, 
detections (open triangles), \MSX\, sources associated with diffuse emission (filled squares) 
and the source identified by Megeath et al. (1996) (open asterisk) and 
Rathborne et al. (2002) (filled asterisk). The numbers mark the sources discussed in 
Sect.~\ref{continuum-sed}.}
\end{center}
\end{figure}

Coincident with a peak in the stellar density distribution is the known cluster Bo~11 and
interestingly, two of the molecular condensations discussed in Sect.~\ref{condensations}. The
source \sIV\, reveals, in \twomass\, images, extended IR emission and a tightly packed group
of sources, many of which are only visible in the K$_{\mathrm s}$-band image (this source will be discussed 
further in later sections).

Many other candidate clusters are identified from this analysis, but display no obvious 
clusters within the \twomass\, images. Many are associated with peaks in the visual extinction 
maps, therefore making the identification of a cluster from the images difficult. While clusters are 
located across the region, it is those associated
with the visual extinction peaks that are potentially the most interesting. This correspondence
suggests the clusters may be embedded within the pillars. Additional
data is required, however, to determine the exact nature of many of these candidate clusters.

\begin{figure*}
\centering
\psfig{file=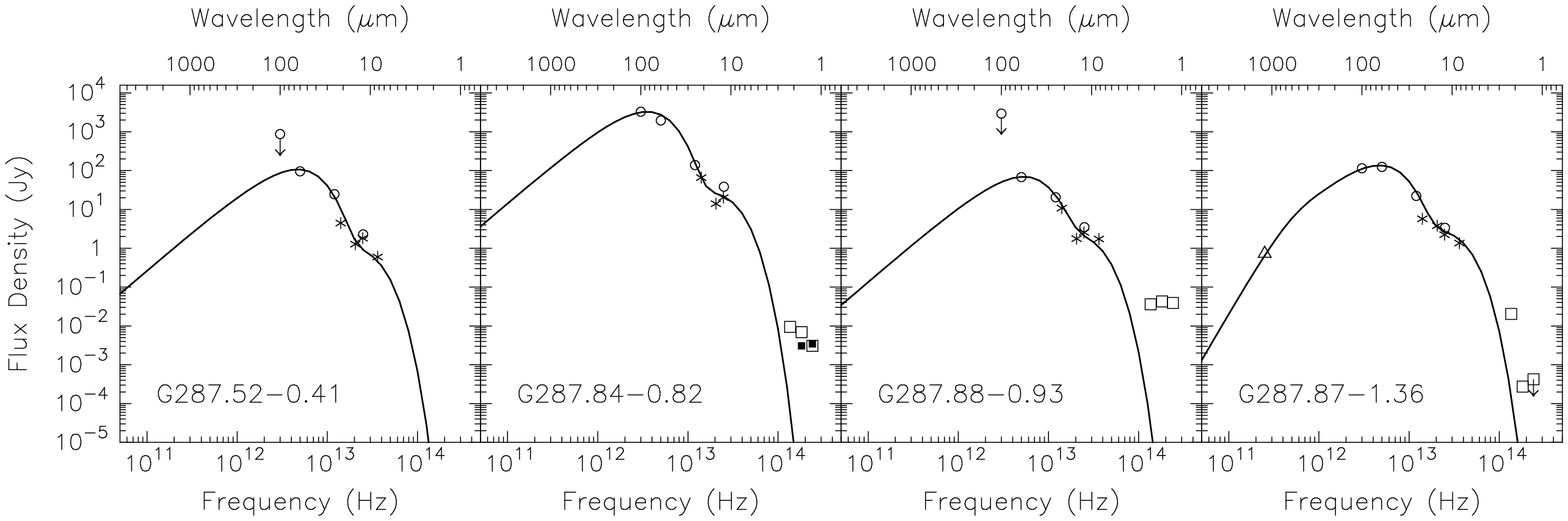,width=0.95\textwidth,angle=-90}
\caption{\label{sed} Spectral energy distributions for the candidate YSOs; 
G287.52-0.41 (source 1), \sIV\, (sources 2), G287.88-0.93 (source 3) and \myso\, (source 4). 
Open squares represent \twomass\, fluxes, stars represent \MSX\, fluxes, while circles mark the
\IRAS\, fluxes. The solid line in each plot is a fit to these points using a modified two component 
black-body function (see text for details). The 8\,\um\, flux is excluded from
the fit for \sIV\, as it is known to be dominated by emission from PAH molecules. The near-IR 
points were not included in the fits as these likely represent emission from an extincted hot
component rather than dust emission.}
\end{figure*}

\begin{table*}
\begin{center}
\caption{\label{sed-properties}The derived parameters of the two-component grey-body fits to the 
SEDs for  G287.52-0.41, \sIV\, G287.88-0.93 and \myso\, as shown in Fig.~\ref{sed}. The 
parameters correspond to the temperatures of the outer (\touter) and inner (\tinner) dust components, the 
inner (\rinner) and outer (\router) radii of the dust shell and estimates of the gas mass (M$_{\mathrm gas}$),
Luminosity derived from these and the corresponding spectral types of the exciting star. The numbers 
in brackets next to the source names refer to the source label from Fig.~\ref{cc-msx}.} 
\begin{tabular}{lccccccc}
\hline
\multicolumn{1}{c}{Source} &\touter &\tinner &\rinner &\router &M$_{\mathrm gas}$$^{a}$ & \multicolumn{1}{c}{Luminosity} & Spectral\\
          &  K      & K     & AU    & AU     & \Msun     & \multicolumn{1}{c}{10${^4}$ \Lsun}   & Type  \\
\hline	                      			      
G287.52-0.41 (1)  & 80   & 320  & 10  & 850   & 16   & 5.1  & O8.5 \\
\sIV\,   (2)      & 60   & 300  & 57  & 7200  & 1200 & 370  & cluster? \\
G287.88-0.93 (3)  & 90   & 330  & 14  & 570   & 7.3  & 3.4  & B0 \\
\myso\, (4)       & 80   & 360  & 14  & 960   & 12   & 3.3  & B0 \\
\hline
\end{tabular}\\
\begin{minipage}{14cm}
$^{a}$\, The mass of \myso\, was determined from the observed flux at 1.2-mm. Estimates of the 1.2-mm
flux for the other sources were obtained from the grey-body fit.\\
\end{minipage}
\end{center}
\end{table*}

\subsection{Identification of candidate YSOs}
\label{myso-candidates}

Recently, \citet{Lumsden02} derived criteria for identifying various galactic plane sources from 
fluxes obtained within the \MSX\, bands. The main focus of their study was to separate massive young 
stellar objects (MYSOs) from sources with similar IR colours, in particular, compact \HII\, regions, 
young compact planetary nebulae, and very dusty evolved stars. From mid-IR colour-colour diagrams, 
they find that young objects satisfy \fea\, $>$ 2, with MYSOs having \fdc\, $>$ 1, while
compact \HII\, regions have \fdc\,$<$ 1. These limits were determined from the location of a
known sample of these objects. The sample of  \HII\, regions originated from sources identified within radio 
continuum surveys \citep{Wood89, Kurtz94, Walsh98}. For the MYSOs, the list was compiled by \citet{Lumsden02},
and consisted of objects with strong emission in the longer \MSX\, bands, and with featureless mid-IR spectra, with
the possible exception of silicate absorption.
In this section, we use these criteria to investigate the nature of the point sources across the Carina Nebula in 
order to determine if, and where, any MYSOs and compact \HII\, regions may exist.

Fig.~\ref{cc-msx} shows the colour-colour diagram for all sources  with
reliable detections in the \MSX\, and/or \IRAS\, PSCs. Also included on this plot is the position of a 
black-body and the limits derived by \citet{Lumsden02} for MYSOs and compact \HII\, regions.

In this colour-colour plane, sources with \fdc\, $\leq$ 1 and \fea\, $\leq$ 2 correspond 
to evolved stars. The remaining sources that satisfy \fea\, $>$ 2, are candidate young objects,
13 of which fall within the compact \HII\, region while 9 are potentially MYSOs. Visual 
inspection of the images reveal the sources displaying an \fea~$>$~10 are associated with
emission and are not point sources (these are marked as filled squares in Fig.~\ref{cc-msx}). 
Thus, their identification as MYSO objects is dubious. Two MYSO candidates remain and
are located in the south of the nebula (labelled on Fig.~\ref{8um-Av} as \myso\, and G287.73-1.01).
These appear isolated from any extended mid-IR, radio continuum emission, or the molecular condensations. 
A candidate compact \HII\, region is, however, associated with one of the molecular condensations (\sIV).

Two sources identified within this analysis have previously been identified as YSOs
 and include a bright-rimmed globule at the edge of the southern molecular cloud 
(marked with an open asterisk; G287.63-0.72; \citeauthor{Megeath96}\,\citeyear{Megeath96}) and 
an embedded O9-star (filled asterisk; G287.47-0.54; 
\citeauthor{Rathborne02}\,\citeyear{Rathborne02}). 

\subsection{Continuum SEDs}
\label{continuum-sed}

To investigate the nature of the YSO candidates, spectral 
energy distributions (SEDs) were constructed. Of the 14 candidates, only 4 of these had reliable and
coincident detections within both the \MSX\, and \IRAS\, PSCs. The first of these sources is 
located in the northern part of the nebula (G287.52-0.41), the second is the source associated
with the molecular condensation (\sIV), the third is located at the tip of a giant
pillar (\sV), while the remaining source is located in the south, and is a MYSO candidate (\myso). These 
are discussed in more detail in Sect.~\ref{individual-sources}. 

Fig.~\ref{sed} shows the continuum SEDs for these sources. Included on these plots are the fluxes 
obtained from the \IRAS\, PSC, as well as those estimated from the \MSX\, and \twomass\, images. 
An additional 1.2\,mm flux is included for the source \myso.
The solid line in each plot is a fit to the data using a grey-body of the form 
B$_{\nu}$(T)(1-exp(-$\tau_{\nu}$))$\Omega$, where B$_{\nu}$(T) is the Planck
function at a temperature T, $\tau_{\nu}$ is the dust optical depth, and $\Omega$ is the solid angle
subtended by the dust shell. The opacity was assumed to have the form $\tau$ = $(\nu/\nu_{0})^{\beta}$
where $\nu_{0}$ is the frequency at which the emission becomes optically thick, with $\beta$ set to
2 (determined from tabulated values of \citeauthor{Ossenkopf94} \citeyear{Ossenkopf94}). A two-component model was 
fit to the data, because a single temperature component was insufficient to describe the observed emission. 
Parameters derived from these fits, including the temperatures and radii of the two dust components,
mass, luminosity, and spectral type of the central star, are listed in Table~\ref{sed-properties}.

Observations at a frequency in which the emission is optically thin allows a determination of the
mass of the material. For the source \myso\, additional data at 1.2-mm were obtained which, when
included in fit, revealed a $\nu_{0}$ of 4.0$\times 10^{11}$~Hz (0.75mm). Thus, the emission at 
1.2\,mm is optically thin and was therefore used to estimate the mass.
The expression {\mbox {$M_{gas} = \frac{S_{\nu}D^{2}}{R_{dg}\kappa_{\nu}B_{\nu}(T_{d})}$}} 
was used, where M$_{gas}$ is the total mass of an isothermal dust source, S$_{\nu}$ is the observed 
flux density at an optically thin frequency $\nu$, $\kappa_{\nu}$ is the mass absorption coefficient, 
R$_{dg}$ is the dust-to-gas mass ratio and B$_{\nu}$(T$_{d}$) is the Planck function at the dust 
temperature \citep{Chini87}. Using values of R$_{dg}$ $\sim$ 0.01, a dust opacity of 0.1 
m$^{2}$/kg (determined from \citeauthor{Ossenkopf94} \citeyear{Ossenkopf94}), and assuming a distance of 2.2~kpc, 
the resulting mass was found to be 12\,\Msun. The mass estimates represent a lower limit, as the
dust opacity, and temperatures are uncertain, and are most likely lower than quoted here.

Luminosity estimates suggest the exciting stars for these sources are OB-stars. The
high mass and luminosity derived for \sIV\, suggest this object may harbour a cluster of stars.

\begin{figure*}
\centering
\begin{tabular}{rrr}
\psfig{file=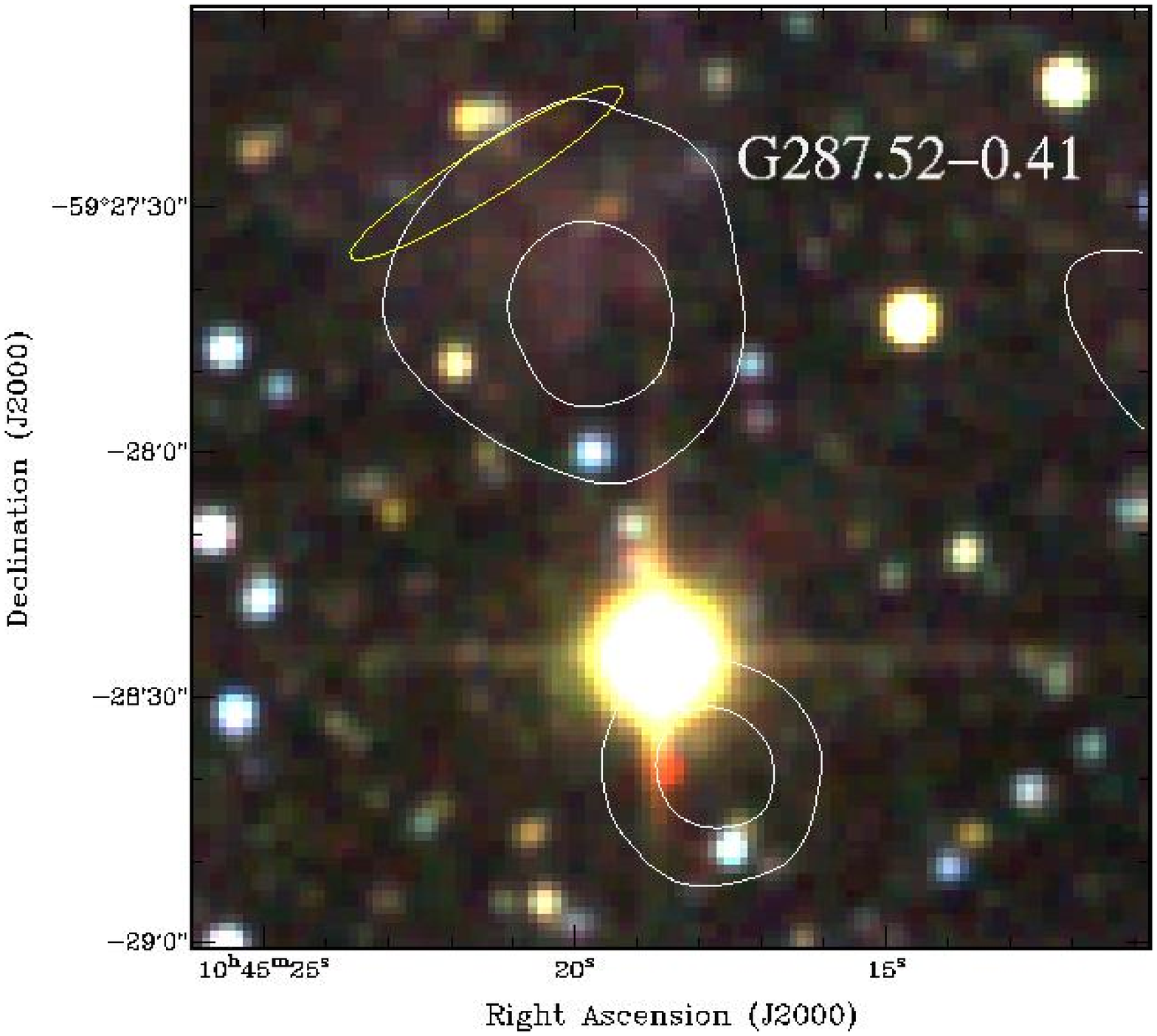,width=0.28\textwidth} &
\psfig{file=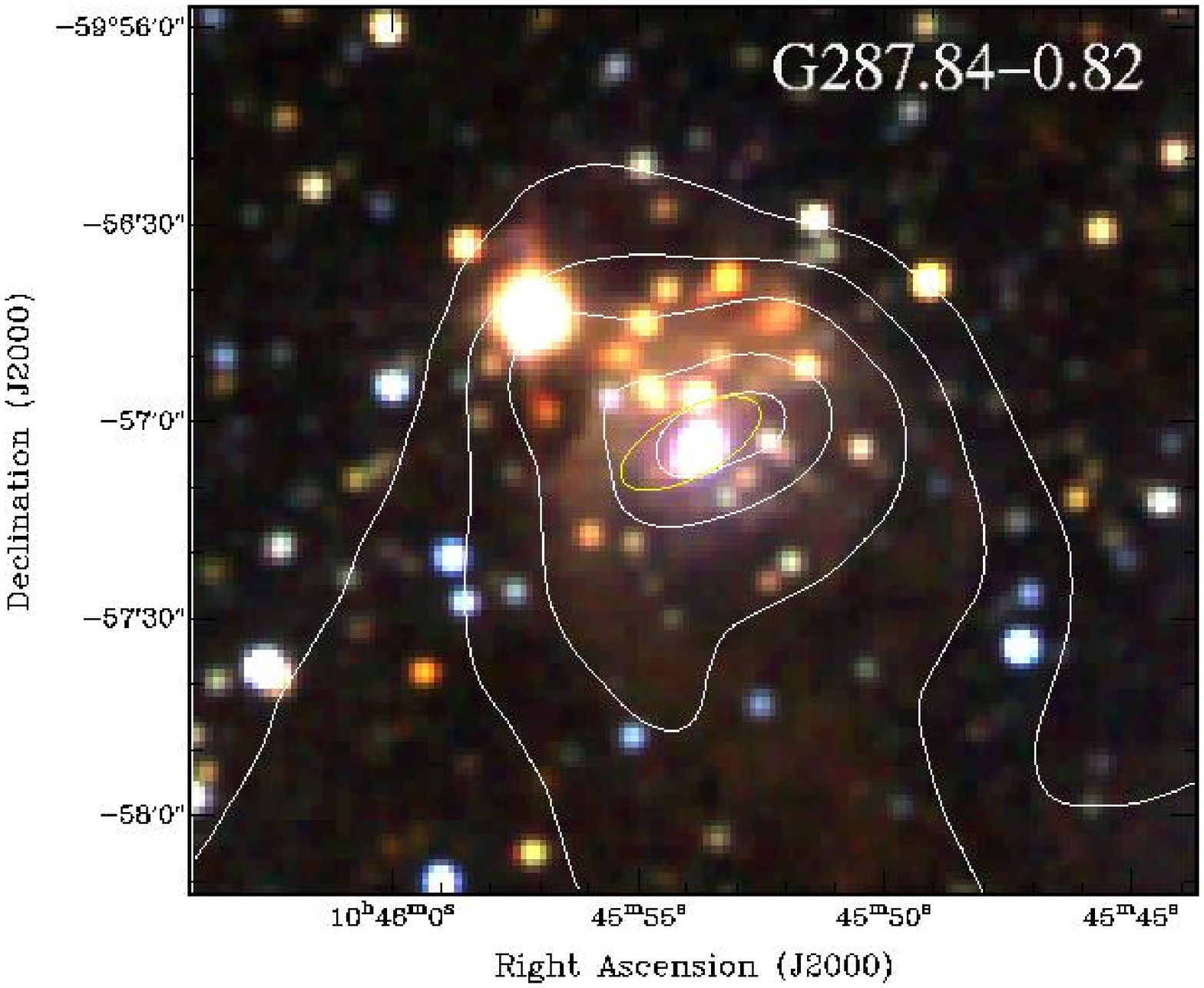,width=0.29\textwidth}&
\psfig{file=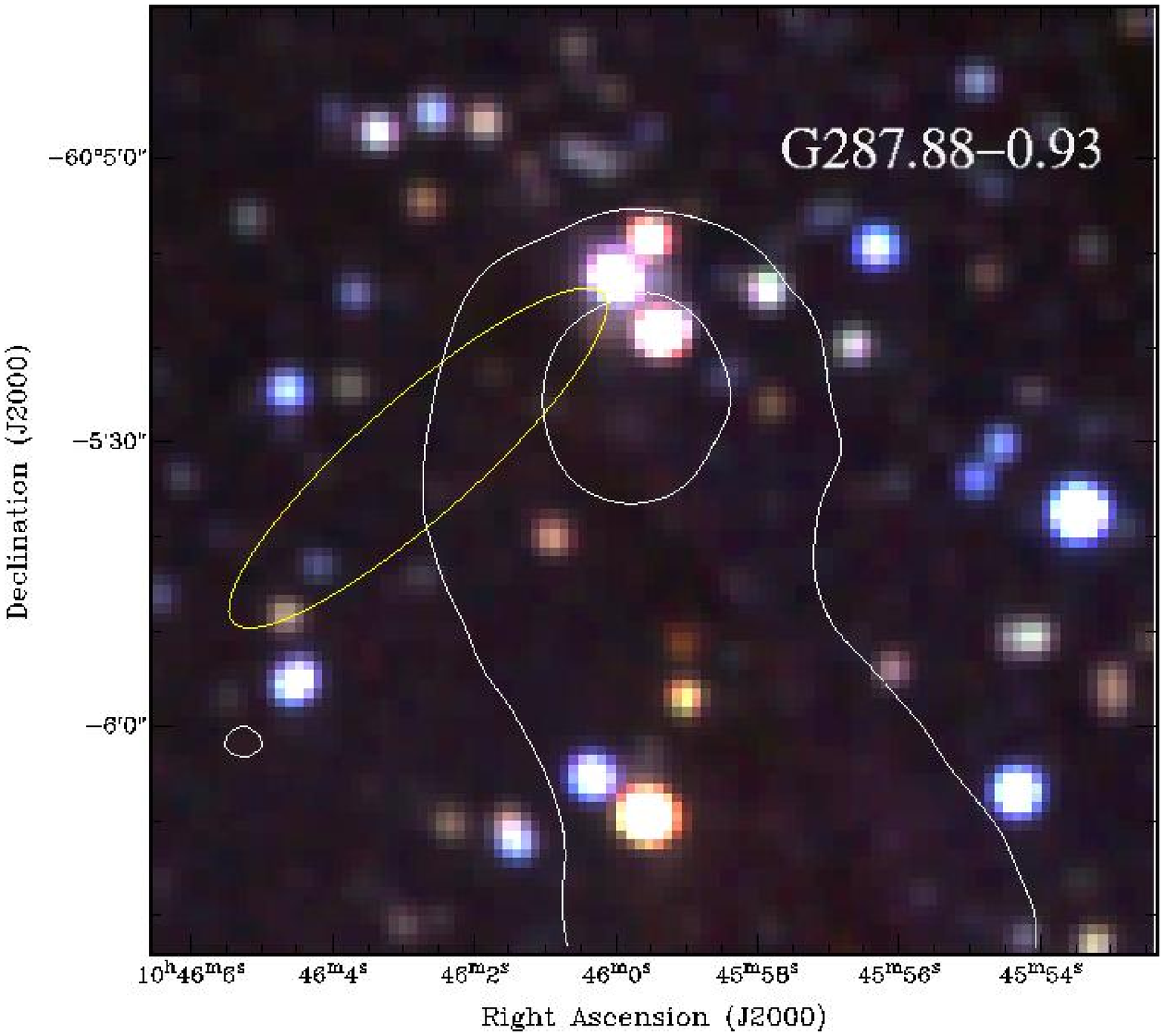,width=0.28\textwidth} \\
\psfig{file=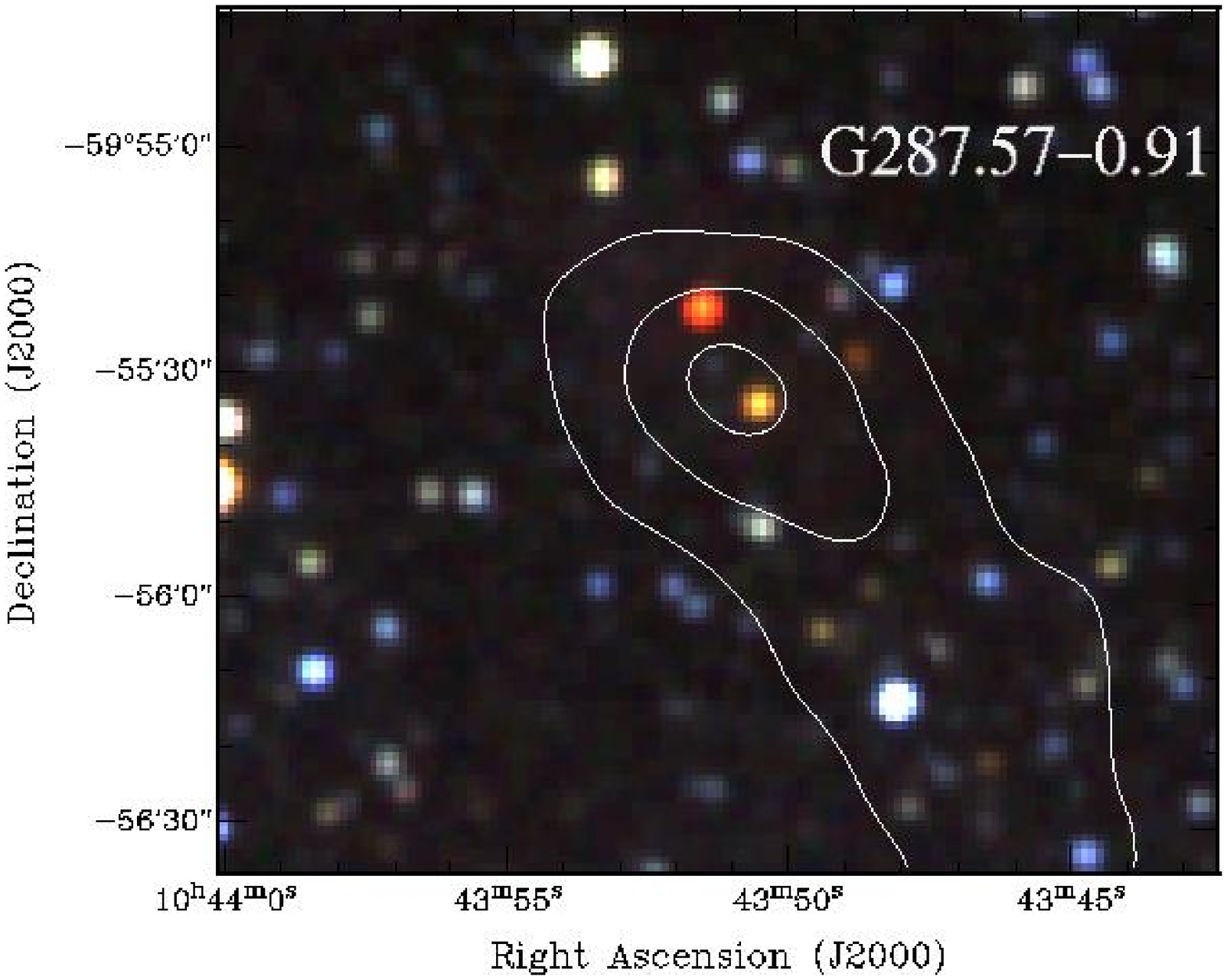,width=0.29\textwidth} &
\psfig{file=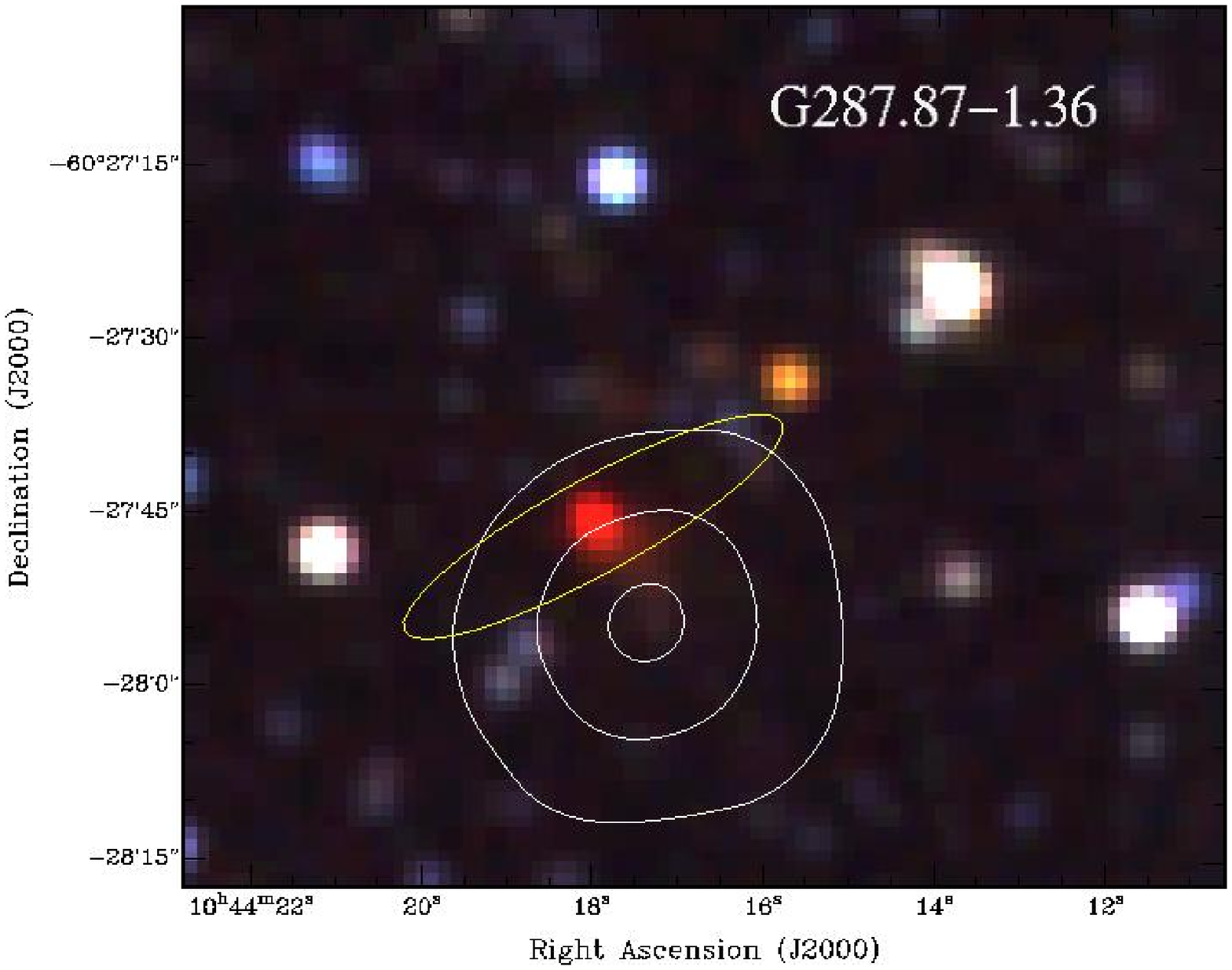,width=0.29\textwidth} &
\psfig{file=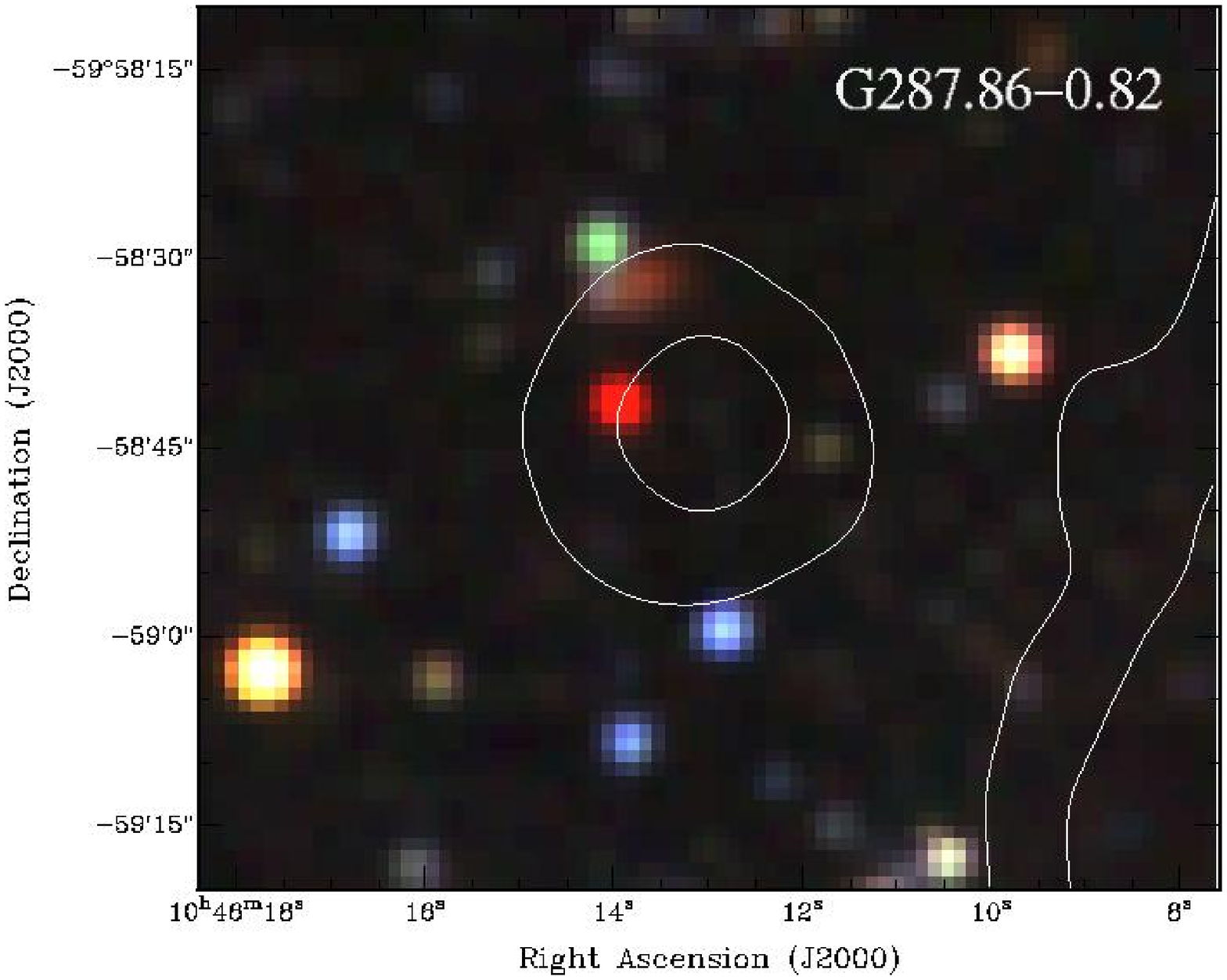,width=0.29\textwidth} \\
\psfig{file=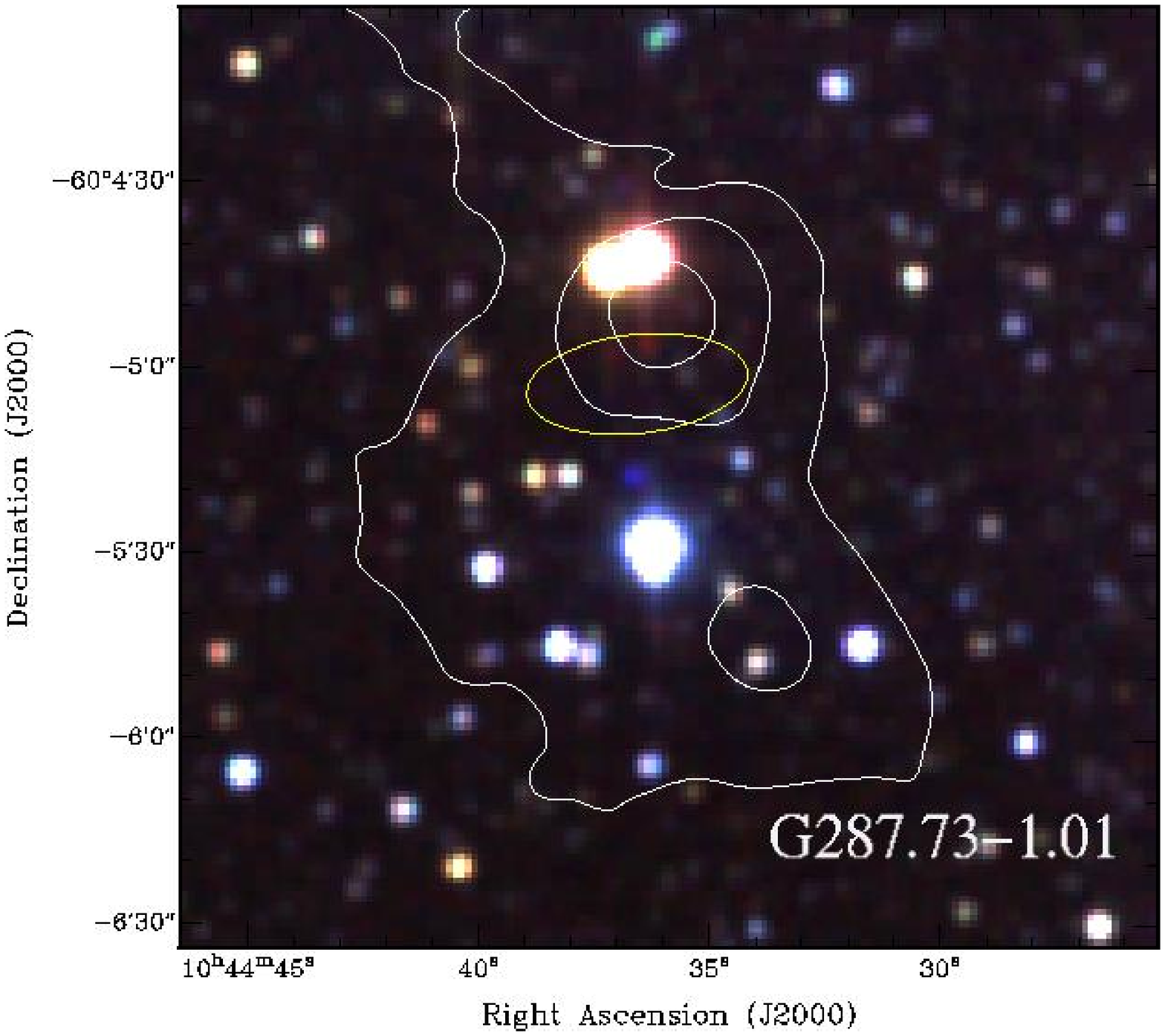,width=0.27\textwidth} &
\psfig{file=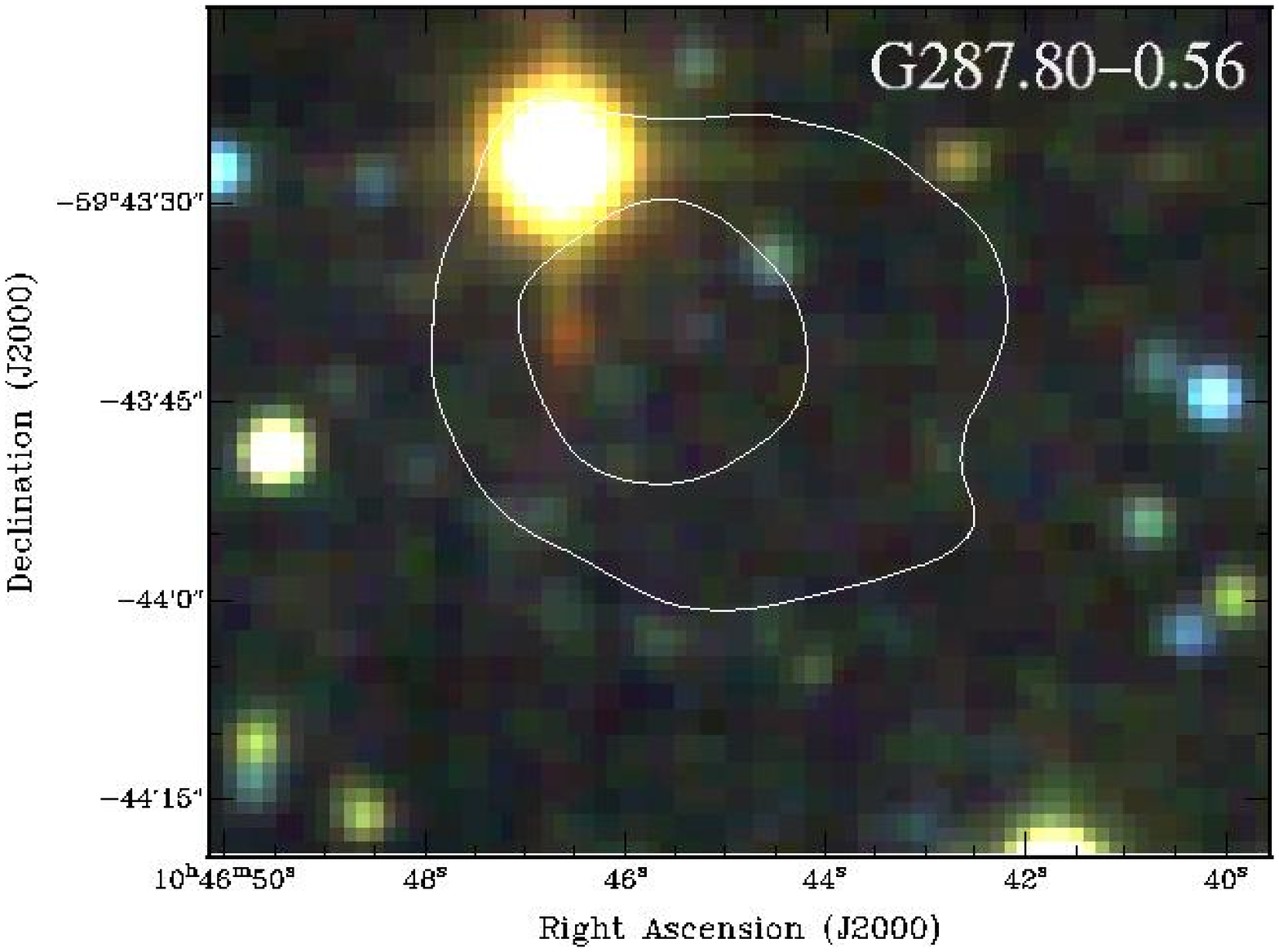,width=0.28\textwidth} &
\psfig{file=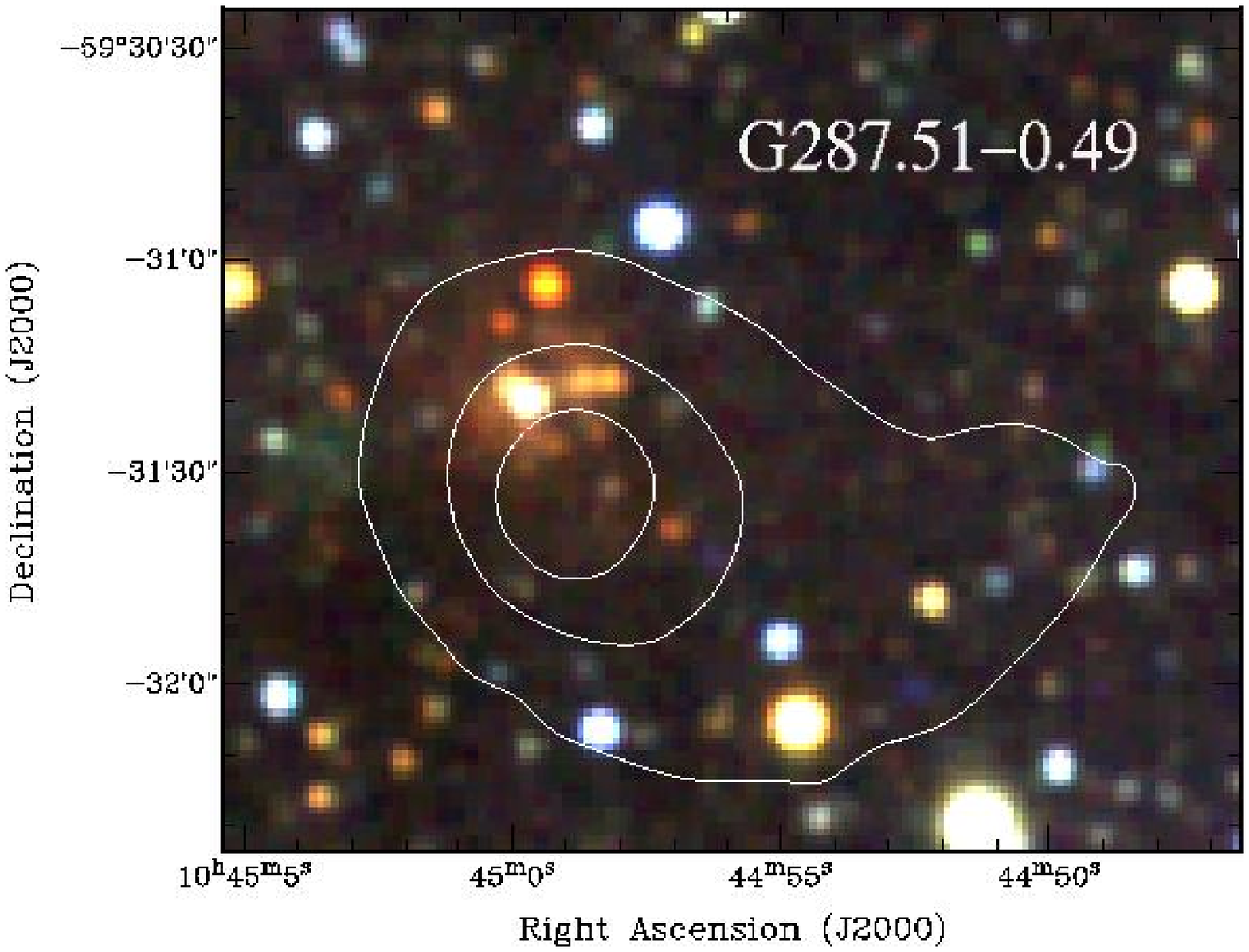,width=0.28\textwidth} \\
\psfig{file=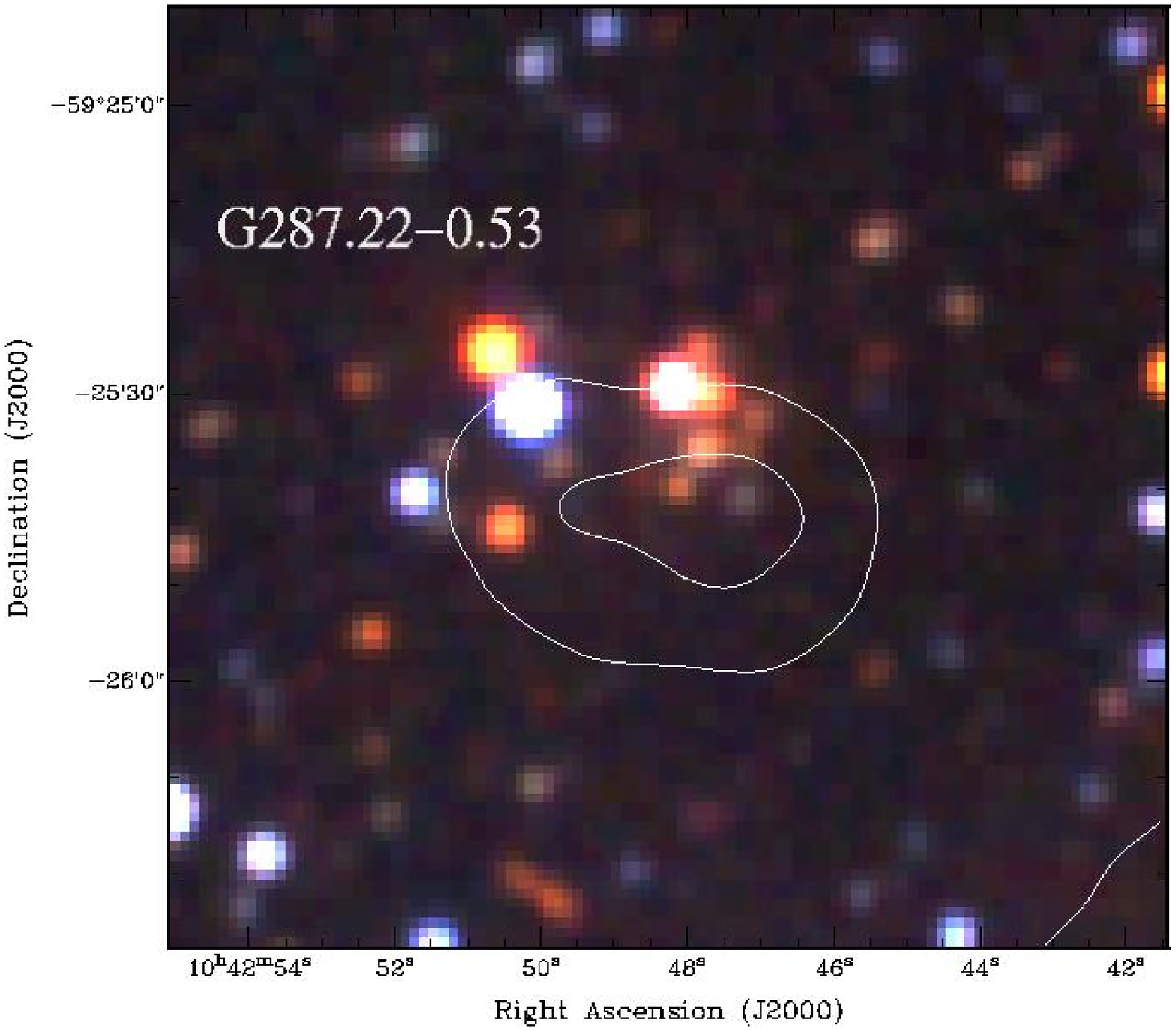,width=0.28\textwidth} &
\psfig{file=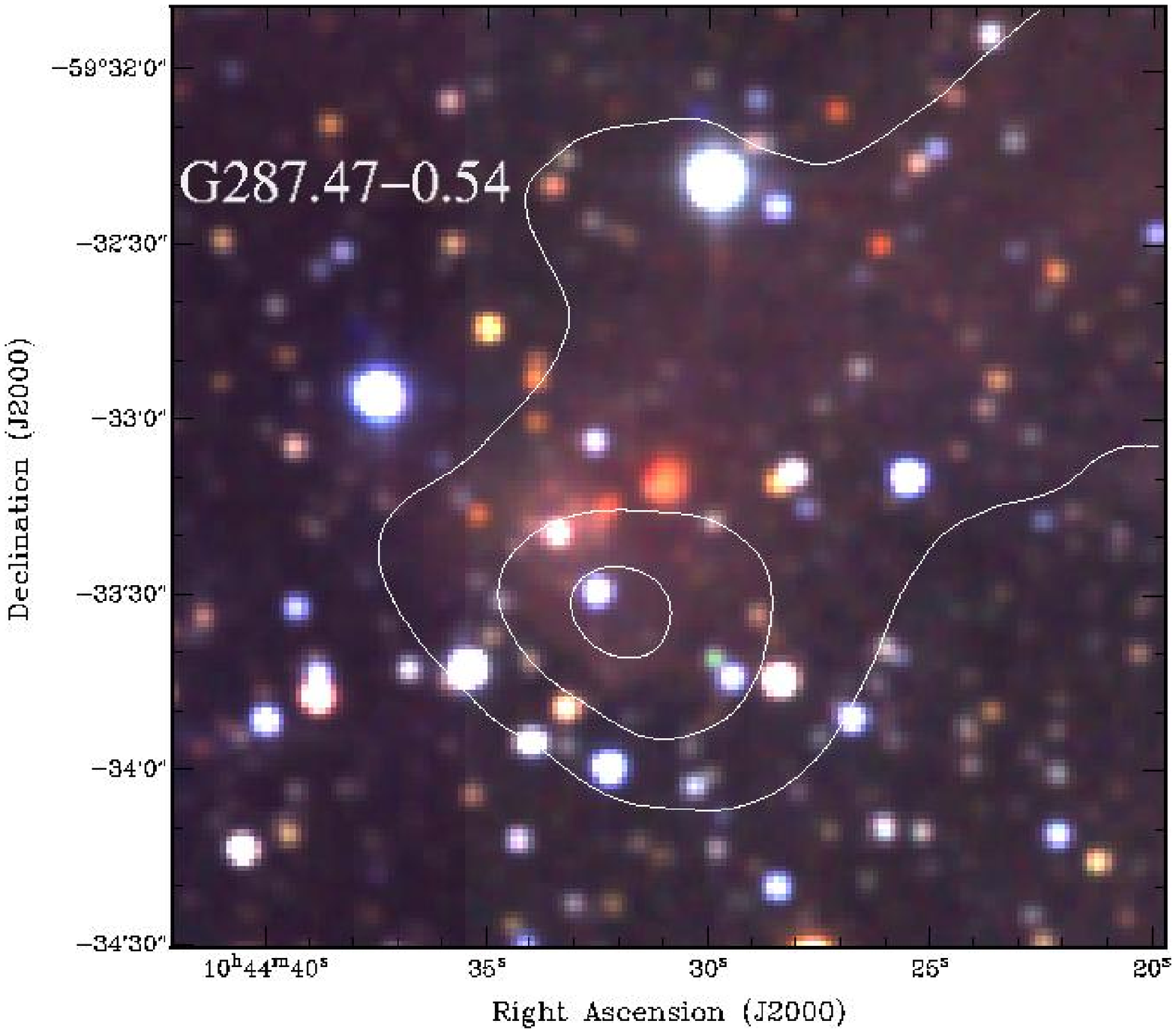,width=0.28\textwidth} &
\psfig{file=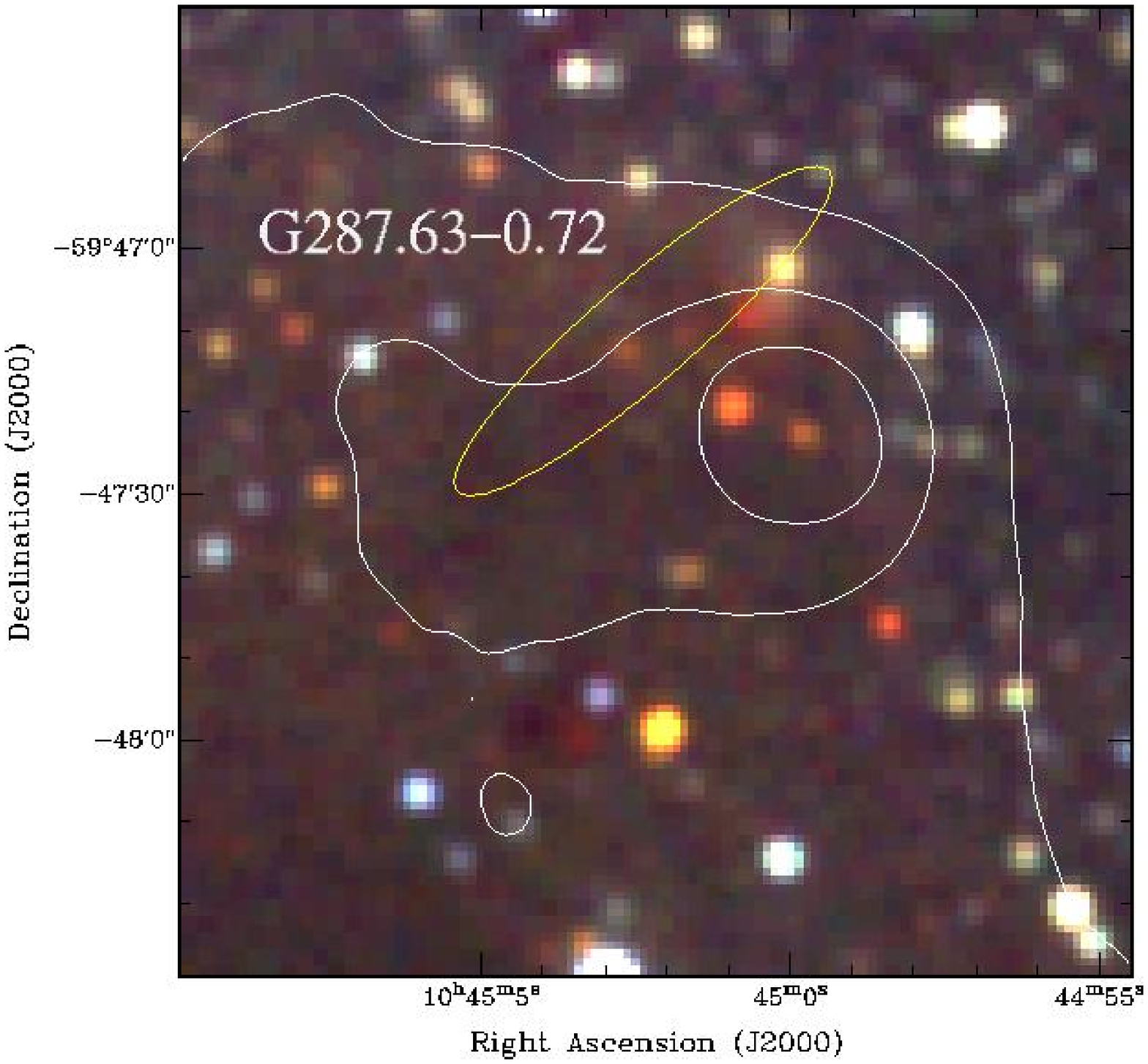,width=0.28\textwidth} \\
\psfig{file=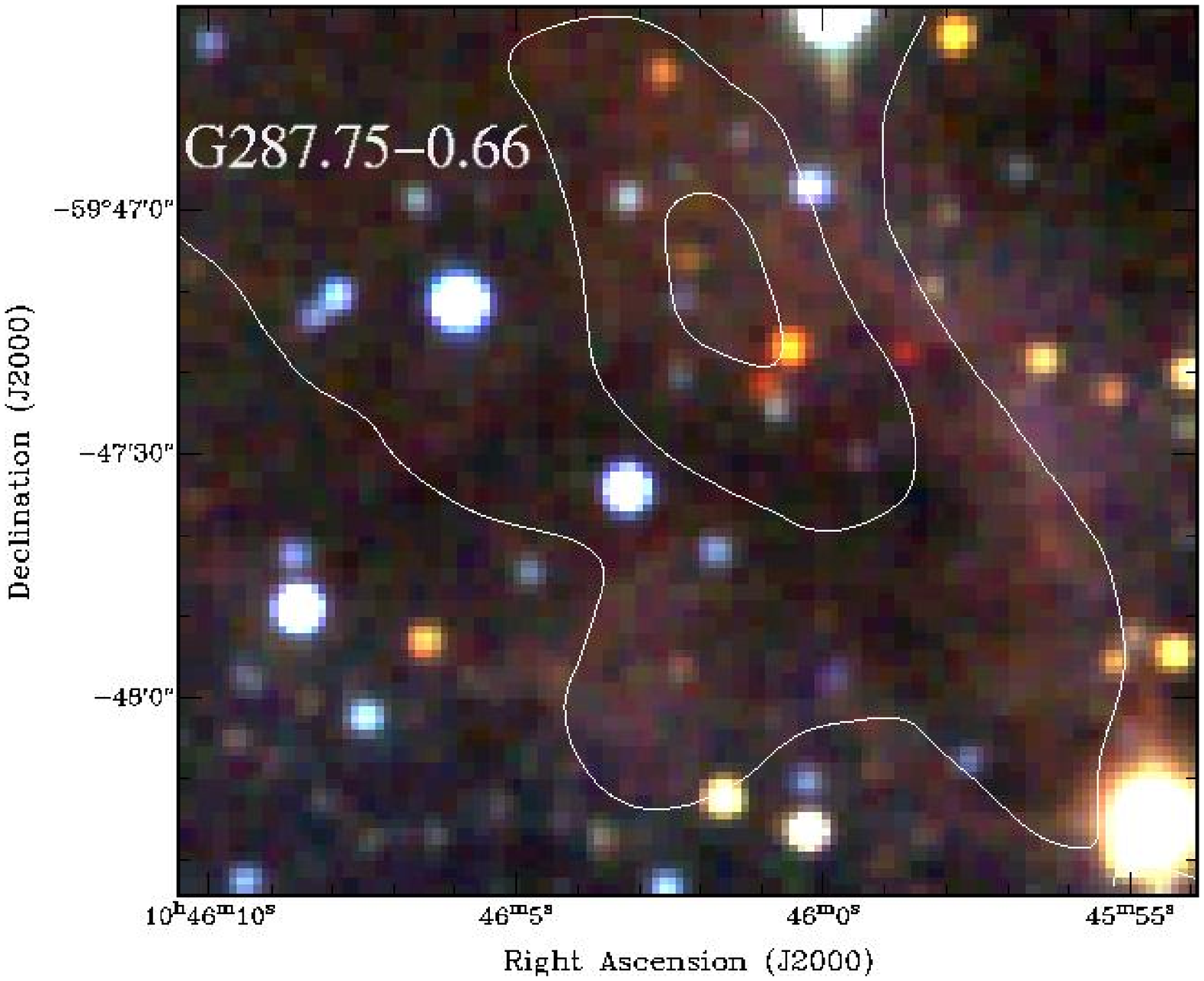,width=0.29\textwidth} &
\psfig{file=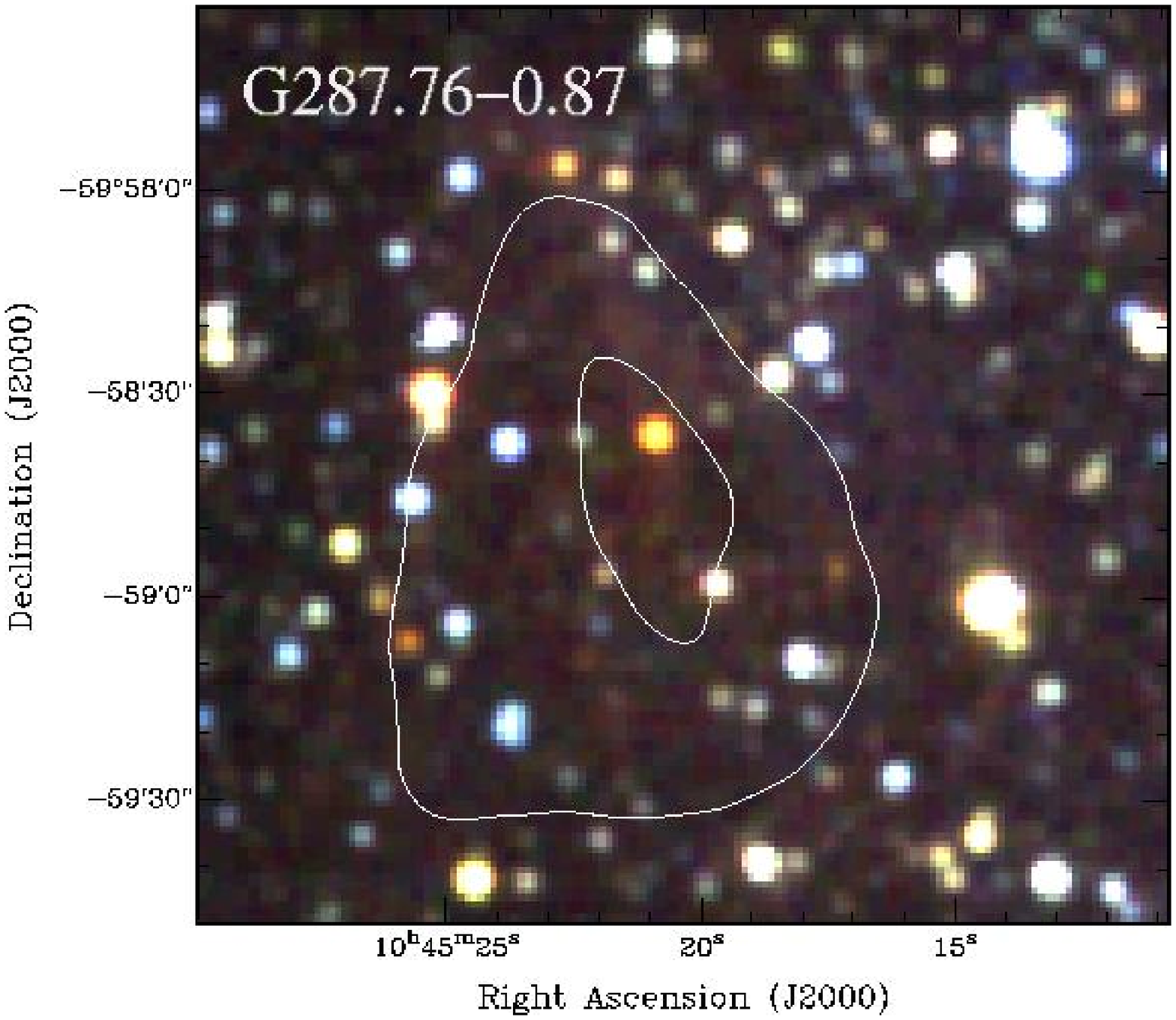,width=0.29\textwidth} &
\psfig{file=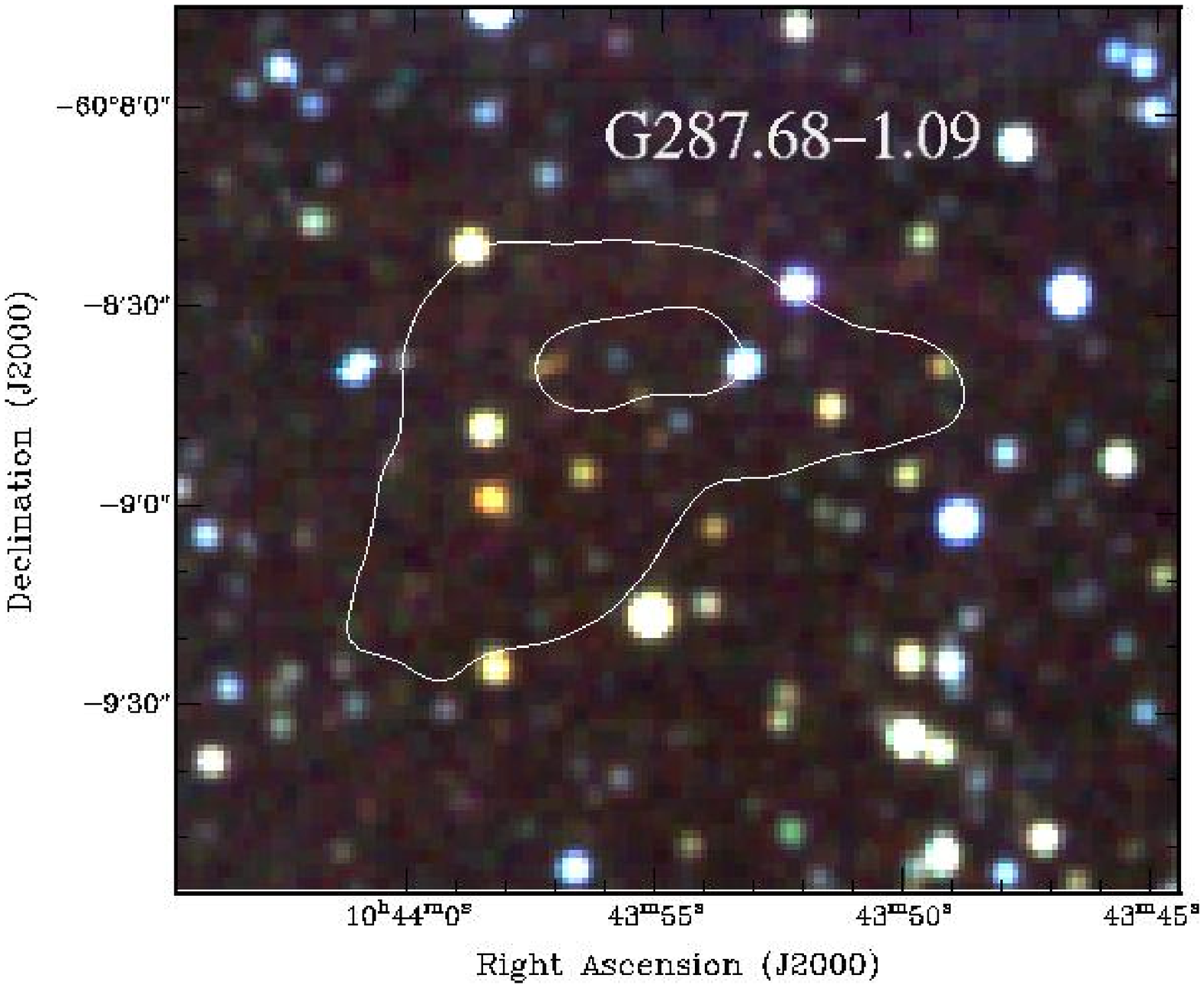,width=0.29\textwidth} \\
\end{tabular}
\caption{\label{candidates}A collection of near-IR three-colour images for candidate YSOs within 
the Carina Nebula (J-band in blue, H-band in green and K$_{\mathrm s}$-band in red). The images are presented
in the order in which they are discussed in Sect.~\ref{individual-sources}. Included also 
are yellow ellipses marking the location of \IRAS\, point sources and contours of 8\,\um\, emission.
Levels are, G287.52-0.41: 0.5, 1; G287.84-0.82: 1, 2, 4, 8, 10; G287.88-0.93: 0.5, 1; G287.57-0.91: 0.5, 
1, 1.5; G287.87-1.36: 0.5, 1, 2; G287.86-0.82: 0.5, 0.7;  G287.73-1.01: 0.5, 1, 2; G287.80-0.56: 1, 2; 
G287.51-0.49: 1, 2, 3; G287.22-0.53: 0.8, 1.3; G287.47-0.54: 2, 4, 6; G287.63-0.72: 1, 2, 3; 
G287.75-0.66: 2, 3, 4; G287.76-0.87: 1, 2; G287.68-1.09: 1, 1.5. All are in 
units of 10$^{-5}$\Wmmsr.}
\end{figure*}

\subsection{Near-IR three-colour images}

Fig.~\ref{candidates} shows near-IR three-colour images constructed from the \twomass\,
data, for the 14 YSO candidates  plus one source of interest (G287.86-0.82). This source was selected
because of its extremely red near-IR colour, similar to the MYSO candidate \myso. These images reveal
any near-IR counterparts to the mid-IR sources, in particular, note the prominant cluster of reddened
stars associated with \sIV. 

Near-IR images of the sources G287.73-1.01 and G287.80-0.56, reveal  bright stellar
objects, the former being associated with a known O7-star (HD 93222; \citeauthor{Walborn72} \citeyear{Walborn72}). 
Although these objects lie in the regime for candidate MYSOs and compact \HII\, regions respectively, they most 
likely represent older stars, as they are clearly seen in the near-IR and, 
for the object G287.73-1.01, also at visible wavelengths. As a result, these are not considered to be candidate YSOs.

Several sources identified as compact \HII\, regions, are associated with diffuse emission in the 
near-IR images. These objects (G287.51-0.49, G287.22-0.53, G287.47-0.54, G287.63-0.72, 
G287.75-0.66, G287.76-0.87 and G287.68-1.09) all show extended K$_{\mathrm s}$-band emission, with several
reddened sources located at the peak of the 8\,\um\, emission.
In addition, all seven of these objects have extended 843\,MHz radio continuum emission associated 
with them.

Four of these objects (G287.51-0.49, G287.22-0.53, G287.47-0.54, and G287.63-0.72) contain
multiple sources and are possibly embedded clusters. These objects were all located external to the stellar density 
map of Sect.~\ref{stellar-density}. The sources
G287.47-0.54 and G287.63-0.72 have been discussed previously in \citet{Rathborne02} and 
\citet{Megeath96}. 

Near-IR images of the remaining three sources, G287.75-0.66, G287.76-0.87 and 
G287.68-1.09, show no obvious embedded stars, with the stellar colours most likely a result of 
interstellar extinction.

\subsection{Synopses of the most interesting YSO candidates}
\label{individual-sources}

\subsubsection{G287.52-0.41}

This source is located in the north of the Carina Nebula and has mid-IR colours indicative of a 
compact \HII\, region (source 1 in Fig.~\ref{cc-msx}). Consistent with this, is the detection of 
843\,MHz radio continuum emission toward this object. Mass and luminosity estimates suggests 
the presence of an embedded O8.5-star. Furthermore, in the near-IR, a single reddened object is
seen and is coincident with the \MSX\, and \IRAS\, point sources.

\subsubsection{The cluster within \sIV}
\label{little-cluster}

This is the only case within the current dataset, for which a clearly 
visible cluster of young stars has been found embedded within a molecular condensation (this 
cluster was first identified by \citeauthor{Dutra01} \citeyear{Dutra01}). It is the 
brightest point source in the mid-IR and 843\,MHz radio continuum maps within the southern Carina Nebula. 
The near-IR three colour image for this cluster, reveals many embedded stars, with 
several detected only at the K$_{\mathrm s}$-band. These sources represent the most 
interesting and youngest within the cluster.

In the mid-IR colour-colour plane, this source (number 2), lies within the region for compact \HII\, 
regions. As expected, it shows strong, compact radio continuum emission. The mass and luminosity 
derived for this source, suggests it contains at least one massive star within a cluster of lower mass stars. 
This is consistent with previous studies which suggest a B1 star and reflection nebula exist here (ALS~1883; 
\citeauthor{Herbst75} \citeyear{Herbst75}). However, 6.7~GHz methanol maser emission was not detected toward this 
object in a survey by \citet{MacLeod98}.

\begin{figure}
\centering
\psfig{file=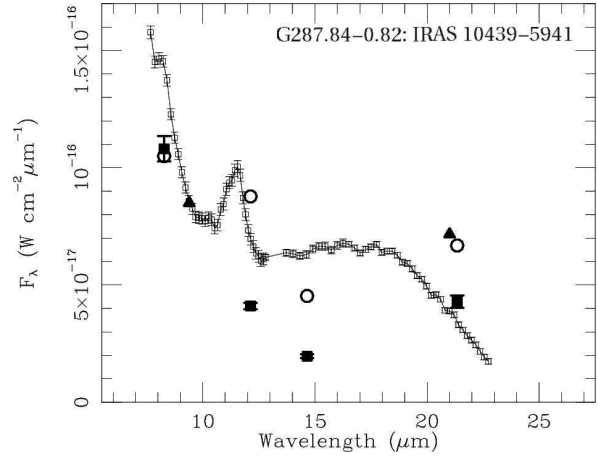,width=0.43\textwidth,angle=-90}
\caption{\label{src6-7lrs}The \IRAS\, LRS spectrum of \sIV\, (open squares) with error bars, and
comparisons with photometry from the \IRAS\, PSC (filled triangles) and \MSX\, PSC (filled squares) 
and from the \MSX\, images (open circles).}
\end{figure}

An averaged \IRAS\, Low Resolution Spectrometer (LRS) spectra was also obtained for this object 
(Fig.~\ref{src6-7lrs}). Within the sizeable LRS 
aperture (5\arcmin\, at 8--13\,\um\, and 7.5\arcmin\, at 11-23\,\um) strong PAH emission 
is seen at 7.7 and 11.3\,\um, confirming our previous identification of PAHs on the basis of the 
\MSX\, and \co\, morphology. Flux densities estimated from the \MSX\, images and PSC, and \IRAS\, PSC, are also 
included on this plot. The obvious discrepancies between flux densities derived from the LRS, 
the \MSX\, image, and those from the \MSX\, PSC, reflect the complexity of the 
region, with its widespread mid-IR emission, and the difficulties in choosing an appropriate 
background level to subtract from the dominant {\mbox {mid-IR}} peak.  

\subsubsection{G287.88-0.93 and G287.57-0.91}
\label{filaments}

The mid-IR morphology of these two sources show filamentary features with point sources at their tips.
They were included within the colour-colour plot of Fig.~\ref{cc-msx} and fall within the regime for compact
\HII\, regions. However, no strong 843\,MHz radio continuum emission is associated with 
them. The source G287.88-0.93 does, however, contain an \IRAS\, point source, with its 
SED corresponding to an embedded B0 star. A star of this spectral type would
produce a compact \HII\, region, albeit weak, and possibly below the detection limits of the MOST
considering the distance to the nebula. 

The near-IR image for G287.88-0.93 shows a group of three stars 
falling within the peak of the 8\,\um\, emission. Near-IR images of G287.57-0.91, reveal a bright K$_{\mathrm s}$-band 
source located at the tip of the filamentary 8\,\um\, feature.

\begin{figure}
\centering
\psfig{file=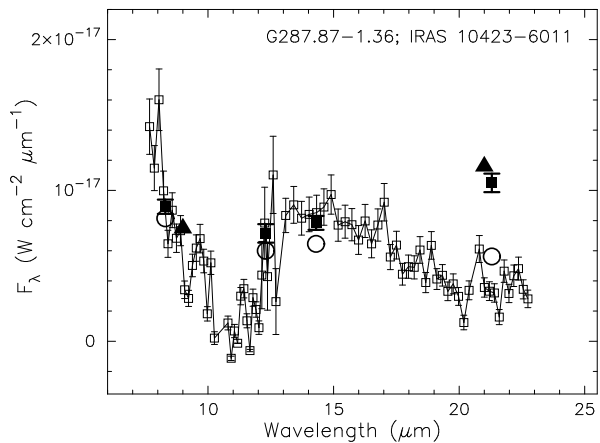,width=0.45\textwidth,angle=-90}
\caption{\label{src16lrs}The \IRAS\, LRS spectrum of \myso\, (open squares) with error bars, and
comparisons with photometry from the \IRAS\, (filled triangles) and \MSX\, (filled squares) PSCs
and from the \MSX\, images (open circles).}
\end{figure}

\subsubsection{\myso}
\label{myso}

This source is the most likely to correspond to a bona-fide MYSO, as it
lies within this regime in the mid-IR colour-colour diagram, and has a corresponding \IRAS\, 
point source. The luminosity derived from its SED suggests an embedded {\mbox {B0-star}}. 
An extremely red object exists in the near-IR three colour image, and is coincident with an unresolved 
mid-IR and 1.2-mm continuum source.
The \IRAS\, LRS  for this source (Fig.~\ref{src16lrs}), clearly shows a deep 
silicate absorption profile at 10\,\um\, indicative of high extinction.

In addition, \citet{Bronfman96} detected CS (2--1) line emission toward this object in their survey 
of \IRAS\, sources having colours of compact \HII\, regions. It appears in their table~1 as 
IRAS~10423-6011, with V$_{lsr}$=13.7\kms, T$_{mb}$=0.84\,K and $\Delta$V=2.7\kms. Because the central 
velocity is blue-shifted by more than 15\kms\, with respect to the other molecular condensations here,
it is not clear if this object is actually associated with the nebula.

\subsubsection{G287.86-0.82}

Located to the east of the cluster \sIV\, is the object G287.86-0.82. This source shows 
weak 8\,\um\, emission, with a strong K$_{\mathrm s}$-band source.  Located nearby, is a second object whose 
emission peaks in the H-band with extended K$_{\mathrm s}$-band emission (green in the near-IR three colour image). Such
emission either arises from a reflection nebula or shocked gas. Detailed
spectra are needed to determine the exact nature of this object.

\subsection{Ongoing star formation within the pillars}

The sources identified here significantly increase the number of young objects found within the Carina 
Nebula. Previous studies located  6 embedded OB-stars in the northern molecular 
cloud complex associated with Tr~14 \citep{Brooks01, Rathborne02}, while a single B0 star is seen
at the edge of the southern molecular cloud \citep{Megeath96}. 

To investigate the age distribution of the star formation within this GMC, the recent rate of 
massive star formation ($\psi$) has been estimated. Studies of the stellar population within the 
clusters Tr~14 and Tr~16 find that the massive stars formed within the past 1--3~Myr \citep{DeGioia01,Tapia03}. 
These clusters have a population of $\sim$ 33 O-type stars, which implies a $\psi$ of 
{\mbox {$\sim$ 1--3$\times$10$^{-5}$ {O-stars yr$^{-1}$}}}. 

Mass and luminosity estimates exist for ten young objects within the Carina Nebula (the seven previously identified 
and the three additional sources identified here). Age estimates were calculations from their 
Kelvin-Helmholtz time, assuming they consist of a single star. Ages were found to range between 
1--5 $\times$10$^{5}$ yrs, which for these 10 sources, suggests $\psi$ to be $\sim$ 3$\times$10$^{-5}$ 
O-stars yr$^{-1}$. This value is consistent with the rate determined from the known optical
O-stars and is expected if the rate of star formation has been constant over the age of the
nebula. While this is possible, these two epochs may instead represent two isolated bursts of star 
formation. Distinguishing between these two mechanisms is difficult with the limited data presented 
here.

The rate of star formation for the Carina Nebula is comparable with estimates for two other well
studied regions of massive star formation. For instance, NGC~3606 contains $\sim$ 50 O-type stars 
which have formed over the last 2--3~Myrs (see \citeauthor{Eisenhauer98} \citeyear{Eisenhauer98} and references 
therein). This 
suggests a $\psi$ of $\sim$ 1--3 $\times$10$^{-5}$ O-stars yr$^{-1}$. While for 30~Dor, at a similar 
age, and containing $\sim$ 100 O-type stars \cite{Walborn97}, the rate is estimated to be 
$\sim$ 3--5 $\times$10$^{-5}$ O-stars yr$^{-1}$.

\section{Conclusions}

We have undertaken a multi-wavelength study incorporating data from \twomass, \MSX, \IRAS, MOST, and 
the SEST, to investigate the nature of, and search for star formation within, the giant pillars of the 
Carina Nebula. Our main results and conclusions are summarised below.

\subsection{Molecular clouds, PDRs and ionization fronts}

Emission from the 8\,\um\, \MSX\, band outlines the known molecular clouds and
giant pillars within the Carina Nebula, and  pinpoints regions where the UV radiation is 
penetrating the molecular material and forming PDRs. Visual extinction maps match extremely well 
the 8\,\um\, emission tracing the dense gas associated with the pillars.

Interestingly, the 21\,\um\, and 843\,MHz radio continuum emission match extremely well and are located along the 
edges of the PDRs closest to the most 
influential clusters. Emission within these bands reveals heated 
$\sim$40 K dust and ionization fronts. The geometry of the largest pillar is consistent with 
interactions from the nearby massive stars carving the molecular material around a 
dense core.

Bright condensations located at the tips of the giant pillars, were found to be externally 
heated, with PDRs located along the edges in the direction of the massive clusters. The properties 
of the molecular material suggest they have the potential to be massive star forming cores. 

\subsection{Evidence for recent star formation activity}

To search for evidence of star formation activity  a stellar number
density map was used. This was derived from the dereddened K$_{\mathrm s}$-band magnitudes for sources in the \twomass\,
PSC. Many candidate young clusters were identified within this map, several of which appear
to be related to, and possibly embedded within, the giant pillars. Interestingly, the brightest IR and
radio continuum source in the region contained the only case of a clearly visible cluster within the
\twomass\, images.

A candidate MYSO and several compact \HII\, regions were identified across the nebula using mid-IR 
colour criteria. SEDs were constructed to study these objects further, and reveal the exciting stars in 
several cases correspond to OB-stars. Table~\ref{coords} lists a summary of all the candidates.

\subsection{Triggered star formation in the pillars?}
The results presented here clearly show that the large molecular clouds within the 
Carina Nebula are being strongly affected by the intense stellar winds and
harsh radiation fields from the nearby massive star clusters. For instance, giant pillars are forming as a result 
of these interactions, as dense cores shield the surrounding material. What still remains unclear 
however, is the effect these interactions have on the ongoing star formation within this GMC. 
Young clusters and candidate MYSOs are found both within, and external to, the giant pillars. 
In addition, while star formating sites are found scattered across the region, there are several locations 
where the intense interactions from the known clusters are potentially triggering the activity. It is also likely
that the rate of star formation has been constant within the nebula since the birth of the massive star 
clusters.

\section*{Acknowledgments}
JMR and MGB acknowledge the support of the Australian Research Council and the University of New 
South Wales. MC thanks NASA for its support of his participation in this work under LTSA grant 
NAG5-7936 with the University of California, Berkeley. This publication makes use of data products 
from 2MASS (a joint project of the UMASS and the IPAC/CIT, funded by NASA and 
the NSF) and \MSX\,(processing of this data was funded by the Ballistic Missile Defense 
Organization with additional support from NASA's Office of Space Science).

\section*{Appendix}

Table~\ref{coords} lists the coordinates and summarizes the properties of the sources discussed within
this work.

\begin{table}[b!]
\begin{center}
\caption{\label{coords}Coordinates for all sources discussed (taken from \MSX\, 
data). Comment descriptions include: massive young stellar object candidate (MYSO); compact \HII\, 
regions (\HII); bright stellar object (S); tip of a filamentary feature (T); molecular condensation 
(MC); cluster (C); diffuse K$_{\mathrm s}$-band emission (DE); red object in \twomass\, images (R) and any 
additional nomenclature. The divisions in the table separate the MYSO, the compact 
\HII\, regions, the sources that most likely correspond to evolved stars,
the source displaying an extremely red colour in \twomass\, images, and the molecular condensations. Fluxes
for these sources are given in \citet{Rathborne-phd}.}
\begin{tabular}{lccl}
\hline
Source & \multicolumn{2}{c}{Coordinates}& \multicolumn{1}{c}{Comments}\\
       & $\alpha_{\footnotesize{J2000}}$ & $\delta_{\footnotesize{J2000}}$&\\
\hline
\myso         & 10 44 17.9&-60 27 46  &MYSO; B0-star\\
              &           &           &IRAS~10423-6011\\
\hline
G287.52-0.41  & 10 45 20.9&-59 27 28  &\HII; O8.5-star\\
              &           &           &IRAS~10434-5911\\
\sIV          &10 45 53.6 & -59 57 10 &\HII; MC; C; \\
              &           &           &OB-cluster;\\
              &           &           &IRAS~10439-5941\\
G287.88-0.93  &10 46 00.9 &-60 05 12  &\HII; T; B0-star\\ 
              &           &           &IRAS~10441-5949\\
G287.57-0.91 & 10 43 51.1 & -59 55 23 & \HII; T\\
G287.51-0.49 & 10 44 59.0 & -59 31 24 & \HII; DE\\
G287.22-0.53 & 10 42 49.2 & -59 25 27 & \HII; DE\\
G287.47-0.54$^{a}$ & 10 44 32.8 & -59:33:20 & \HII; DE\\
G287.63-0.72$^{b}$ & 10 45 01.1 & -59 47 06 &\HII; DE\\
	           &            &           &IRAS~10430-5931\\ 
G287.75-0.66 & 10 46 01.2 & -59 46 58 & \HII; DE\\
G287.76-0.87 & 10 45 22.3 & -59 58 23 & \HII\\
G287.68-1.09 & 10 43 57.2 & -60 08 25 & \HII\\
\hline
G287.73-1.01 & 10 44 34.6 & -60 05 36 & S; HD 93222\\
G287.80-0.56 & 10 46 46.8 & -59 43 25 & S \\
\hline
G287.86-0.82 & 10 46 13.9 & -59 58 41 & R\\
\hline
\sIII                &10 44 45.4&-59 59 16& MC\\
\sV                  &10 45 56.1&-60 08 50& MC; C?\\
\sVI                 &10 47 35.4&-60 02 51& MC\\
\hline
\end{tabular}
\end{center}
$^{a}$ Source N4 from \citet{Rathborne02}\\
$^{b}$ Identified by \citet{Megeath96}\\
\end{table}

\end{document}